\documentclass[twocolumn,epjc3]{svjour3}
\usepackage{graphics}
\usepackage{microtype}

\usepackage[hyperindex=true,colorlinks=true]{hyperref}
 
\hypersetup{
    citecolor=blue,
    linkcolor=blue,
    urlcolor=blue
} 

\usepackage{url}
\usepackage{doi}
\usepackage{slashed}
\usepackage{physics}
\usepackage{feynmp-auto}
\usepackage{multirow}
\usepackage{enumitem}
\usepackage{amsmath}
\usepackage{amsfonts}
\usepackage{amssymb}
\usepackage{bm}
\usepackage{mathrsfs}

\usepackage{epsfig}

\usepackage{txfonts}

\usepackage{comment}
\usepackage{subfigure}
\usepackage{multicol}

\usepackage[sort&compress,numbers]{natbib}

\newcommand{\ee}[1]{e_{\,\,\!_{#1}}^{\,\,\!_{\text{IT}}}}

\def\tt{\hat \tau}

\def\lnl{$NN$+$3N\text{(lnl)}$}
\def\sat{NNLO$_{\text{sat}}$}

\allowdisplaybreaks

\begin{document}

\title{Importance truncation in non-perturbative many-body techniques}
\subtitle{Gorkov self-consistent Green's function calculations}




\author{
A.~Porro\thanksref{ad:irfu,em:ap} \and 
V.~Som\`a\thanksref{ad:irfu,em:vs} \and
A.~Tichai\thanksref{ad:tud,ad:emmi,ad:mpik,em:at} \and 
T.~Duguet\thanksref{ad:irfu,ad:leuven,em:td}
}
\date{Received: \today{} / Accepted: date}

\thankstext{em:ap}{\email{andrea.porro@cea.fr}}
\thankstext{em:vs}{\email{vittorio.soma@cea.fr}}
\thankstext{em:at}{\email{alexander.tichai@physik.tu-darmstadt.de}}
\thankstext{em:td}{\email{thomas.duguet@cea.fr}}

\institute{%
\label{ad:irfu}%
IRFU, CEA, Universit\'e Paris-Saclay, 91191 Gif-sur-Yvette, France
\and
\label{ad:tud}%
Technische Universit\"at Darmstadt, Department of Physics, 64289 Darmstadt, Germany
\and 
\label{ad:emmi}%
ExtreMe Matter Institute EMMI, GSI Helmholtzzentrum f\"ur Schwerionenforschung GmbH, 64291 Darmstadt, Germany
\and 
\label{ad:mpik}%
Max-Planck-Institut f\"ur Kernphysik, Saupfercheckweg 1, 69117 Heidelberg, Germany
\and
\label{ad:leuven}%
KU Leuven, Instituut voor Kern- en Stralingsfysica, 3001 Leuven, Belgium
}

\maketitle

\begin{abstract}
Expansion many-body methods correspond to solving complex tensor networks. The (iterative) solving of the network and the (repeated) storage of the unknown tensors requires a computing power growing polynomially with the size of basis of the one-body Hilbert space one is working with. Thanks to current computer capabilities, \textit{ab initio} calculations of nuclei up to mass $A\sim100$ delivering a few percent accuracy are routinely feasible today. However, the runtime and memory costs become quickly prohibitive as one attempts (possibly at the same time) (i) to reach out to heavier nuclei, (ii) to employ symmetry-breaking reference states to access open-shell nuclei and (iii) to aim for yet a greater accuracy. The challenge is particularly exacerbated for non-perturbative methods involving the repeated storage of (high-rank) tensors obtained via iterative solutions of non-linear equations. 
The present work addresses the formal and numerical implementations of so-called importance truncation (IT) techniques within the frame of one particular non-perturbative expansion method, i.e., Gorkov Self-Consistent Green's Function (GSCGF) theory, with the goal to eventually overcome above-mentioned limitations. 
By {\it a priori} truncating irrelevant tensor entries, IT techniques are shown to reduce the storage  to less than 1\% of its original cost in realistic GSCGF calculations performed at the ADC(2) level while maintaining a 1\% accuracy on the correlation energy. 
%
The future steps will be to extend the present development to the next, e.g., ADC(3), truncation level and to SCGF calculations applicable to doubly open-shell nuclei.
\end{abstract}

\section{Introduction}
\label{sec_intro}

Many-body methods applicable beyond the lightest nuclei are essentially based on a systematic expansion of the exact solution of the Schr\"odinger equation around a conveniently chosen reference state. These polynomially scaling many-body techniques are denoted as \textit{expansion methods} and typically sum dynamical correlations via elementary excitations of increasing rank on top of the reference state, either perturbatively or non-perturbatively. Examples of such methods are many-body perturbation theory (MBPT)~\cite{Langhammer:2012jx,Tichai2016,Hu:2016txm,Tichai:2017rqe,Tichai:2018mll,Arthuis:2018yoo,Tichai:2020dna}, self-consistent Green's function (SCGF)~\cite{Dickhoff:2004xx,Soma:2011aj,Carbone:2013eqa,Soma:2013xha,Lapoux:2016exf,Duguet:2016wwr,Raimondi:2018mtv,Arthuis20,Soma20b}, coupled-cluster (CC)~\cite{Hagen:2012sh,Binder:2013oea,Binder:2013xaa,Hagen:2013nca,Signoracci:2014dia,Morris:2017vxi} or the in-medium similarity renormalization group (IMSRG)~\cite{Tsukiyama2011prl,Hergert:2015awm,Hergert:2016iju,Parzuchowski:2017wcq,Morris:2017vxi,Heinz2021} approaches.  To extend their reach to open-shell nuclei and capture noticeably challenging {\it static} correlations,  single-reference methods have been generalized to employ symmetry-breaking reference states~\cite{Soma:2011aj,Signoracci:2014dia,Tichai:2018mll} and to the possible restoration of the broken symmetry in a two-step approach~\cite{Duguet:2014jja,Duguet:2015yle,qiu17a,Qiu:2018edx}. 

Even if in many cases MBPT yields a satisfactory description of nuclear systems (see Ref.~\cite{Tichai:2020dna} for a recent review), it is relevant to push the development of methods that can solve the Schr\"odinger equation to all orders in perturbation theory. In non-perturbative frameworks, all-order (partial) resummations of MBPT contributions are indeed accounted for by solving a set of non-linear equations, e.g.,  Dyson's (Gorkov's) equation in Dyson (Gorkov) SCGF theory. 

In all expansion methods, the working equations are eventually expressed as tensor networks. The involved tensors relate both to the many-body Hamiltonian and to the unknown amplitudes at play in the particular expansion method of interest. Ultimately, the memory load and the computational cost of a given many-body implementation are respectively driven by the highest-mode tensor carried in the calculation and by the complexity of the tensor network associated with the working equations at a given truncation order. In non-perturbative approaches, as SCGF theory, this cost is significantly augmented by the iterative character of the method that requires a repeated computation and storage of the unknown tensors. This issue embodies the bottleneck making difficult to extend nuclear \textit{ab initio} methods to greater accuracy or to very large bases, the latter being needed to reach heavy nuclei and/or to break SU(2) symmetry (possibly on top of U(1) symmetry) to access doubly open-shell nuclei. 

Consequently, expanding the reach and accuracy of \textit{ab initio} calculations can be significantly helped by reducing the computational cost of the non-perturbative method at play. In this context, importance truncation (IT) techniques may play a crucial role by reducing the size of the many-body tensors involved in the calculation, hence the computational cost of the method~\cite{Roth2009, roth2007ab, tichai2019pre}. IT techniques aim at evaluating \textit{a priori} the importance of tensor entries by means of an inexpensive method (e.g., MBPT) and at taking into account only the most relevant ones in the non-perturbative calculation, eventually accounting perturbatively for the discarded entries. In this way the size of the problem is effectively reduced and larger spaces or higher truncation orders, otherwise out of reach, can become accessible to many-body practitioners. The present work aims at formalising, implementing and testing IT techniques within the SCGF framework.  

The article is organized as follows. Section~\ref{sec_GSCGF} introduces the technical aspects of Gorkov SCGF (GSCGF) formalism that are directly relevant to the present work\footnote{For a more complete account of the method, the reader is referred to Refs.~\cite{Soma:2011aj,Soma:2013ona}.}. Section~\ref{sec_measures} discusses the construction of several IT measures suited to the GSCGF formalism. Such measures are afterwards tested in numerical applications, whose results are presented in Sec.~\ref{sec_results}. Conclusions are eventually provided in Sec.~\ref{sec_conclusions}. An appendix details the derivation of one of the IT measures introduced in Sec.~\ref{sec_measures}.

\section{Gorkov SCGF many-body formalism}
\label{sec_GSCGF}

\subsection{Basics}
\label{sec_basics}

In an arbitrary basis  $\{c_p^\dagger\}$  of the one-body Hilbert space ${\cal H}_1$, the many-body Hamiltonian reads in second quantised form as
\begin{align}
H &\equiv T + V + W +... \nonumber \\
&\equiv  \frac{1}{(1!)^2} \sum _{pq} t_{pq} c^{\dagger}_{p} c_{q} \nonumber \\ 
&+ \frac{1}{(2!)^2} \sum _{pqrs} \bar{v}_{pqrs}  c^{\dagger}_{p} c^{\dagger}_{q} c_{s} c_{r}   \nonumber \\
&+ \frac{1}{(3!)^2} \sum_{pqrstu} \bar{w}_{pqrstu} c^{\dagger}_{p}c^{\dagger}_{q}c^{\dagger}_{r}c_{u}c_{t}c_{s}+... \, , 
\label{eq:ham} 
\end{align}
where $T$ denotes the kinetic energy whereas $V$ and $W$ characterize two- and three-body interactions, respectively. Dots embody possible higher-body operators. The three-body interaction $W$ is typically treated at the effective two-body level~\cite{Carbone:2013eqa,Ripoche:2019nmy,Frosini:2021tuj} such that the many-body formalism is specified on the basis of a sole two-body interaction operator $V$.

Starting from basis $\{c_a^\dagger\}$, it is convenient to introduce the partner basis $\{\bar{c}_a^\dagger\}$ defined through
\begin{equation}
	\bar{c}_a^\dagger\equiv\eta_a c_{\tilde{a}}^\dagger\,,\quad\quad\bar{c}_a\equiv\eta_a c_{\tilde{a}}\,,
\end{equation}
which corresponds to exchanging each one-body state $a$ with its time-reversed partner $\tilde{a}$ up to a phase $\eta_a$. By convention $\tilde{\tilde{a}}=a$, with $\eta_a\eta_{\tilde{a}}=-1$.  With both bases at hand, Nambu's representation introduces the two-component vectors
\begin{subequations}
\label{eq:gen_op}
\begin{align}
		\textbf{C}_a &\equiv\begin{bmatrix}
		c_a\\
		\bar{c}_a^{\dagger}
		\end{bmatrix}\,, \\
		\textbf{C}^{\dagger}_a &=\begin{bmatrix}
		c_a^{\dagger}&\bar{c}_a
		\end{bmatrix}\,,
\end{align}
\end{subequations}
fulfilling anti-commutation relations
\begin{equation}
	\bigl\{C_a^{g_1},C_b^{g_2\dagger}\bigr\}=\delta_{g_1g_2}\delta_{ab}\,,
\end{equation}
where $g_1=1,2$ ($g_2=1,2$) labels the rows (columns) of the annihilation (creation) operator. 

Gorkov SCGF theory first introduces a $2\times 2$ matrix of one-body Green's functions
\begin{align}
	i\bm{\mathcal{G}}_{ab}(t,t') 
	&\equiv\bra{\Psi_0}T[\textbf{C}_a(t)\textbf{C}_b^{\dagger}(t')]\ket{\Psi_0} \notag  \\
	&=i\begin{bmatrix}
	\mathcal{G}_{ab}^{11}(t,t')&\mathcal{G}_{ab}^{12}(t,t')\\
	\mathcal{G}_{ab}^{21}(t,t')&\mathcal{G}_{ab}^{22}(t,t')
	\end{bmatrix}\,,
	\label{eq:nambu_prop}
\end{align}
with $T$ the time-ordering operator and where $\ket{\Psi_0}$ denotes the exact ground state of the system. The propagator $\mathcal{G}_{ab}^{g_1g_2}$ is said to be \textit{normal} (\textit{anomalous})  when the two Nambu indices $(g_1,g_2)$ are identical (different).  Time-dependent creation and annihilation operators evolve according to the Heisenberg representation 
\begin{subequations}
	\begin{align}
		c_a(t)&\equiv e^{i\Omega t}c_ae^{-i\Omega t}\,,\\
	c_a^\dagger(t)&\equiv e^{i\Omega t}c_a^\dagger e^{-i\Omega t}\,,
	\end{align}
\end{subequations}
where $\Omega \equiv H - \lambda A$ denotes the grand potential combining the Hamiltonian with a Lagrange term associated with the control of the average particle number\footnote{In practice, two different Lagrange parameters are needed to constrain proton and neutron numbers independently, i.e. $\lambda A$ actually stands for $\lambda_Z Z + \lambda_N N$.}. This feature relates to the fact that GSCGF theory eventually expands the exact ground-state around a particle-number breaking Bogoliubov reference state.

Introducing the spectroscopic amplitudes
\begin{subequations}
	\label{eq:gor_spec_ampl}
\begin{align}
		\textbf{X}_a^{k\dagger}&\equiv\bra{\Psi_k}\textbf{C}_a^\dagger\ket{\Psi_0}=\begin{bmatrix}
		\mathcal{U}_a^{k*}&\mathcal{V}_a^{k*}
		\end{bmatrix}\,,\\
	\textbf{Y}_a^{k\dagger}&\equiv\bra{\Psi_k}\textbf{C}_a\ket{\Psi_0}=\begin{bmatrix}
	\bar{\mathcal{U}}_a^{k*}\\\bar{\mathcal{V}}_a^{k*}
	\end{bmatrix}\,,
\end{align}
\end{subequations}
the Lehmann representation of the propagator in the energy domain 
\begin{equation}
	\bm{\mathcal{G}}_{ab}(t,t')\equiv \int\frac{d\omega}{2\pi}e^{-i\omega(t-t')}\;\bm{\mathcal{G}}_{ab}(\omega)
\end{equation}
is obtained as
\begin{equation}
	\bm{\mathcal{G}}_{ab}(\omega)=\sum_k\biggl\{\frac{\textbf{X}_a^k\textbf{X}_b^{k\dagger}}{\omega-\omega_k+i\eta}+\frac{\textbf{Y}_a^k\textbf{Y}_b^{k\dagger}}{\omega+\omega_k-i\eta}\biggr\}\,. \label{Gorkov_prog_Lehmann}
\end{equation}
The poles of the propagator are given by $\omega_k\equiv\Omega_k-\Omega_0$ with $\{\Omega_k\}$ the eigenvalues of $\Omega$ over Fock space. 

Introducing a partitioning of the grand potential 
\begin{equation}
	\Omega = \Omega_0 + \Omega_1
\end{equation}
such that $\Omega_0$ admits the Bogoliubov vacuum $| \Phi_0 \rangle$ as its ground state, the propagator can be shown to satisfy the implicit Gorkov equation  
\begin{equation}
	\bm{\mathcal{G}}_{ab}(\omega)=\bm{\mathcal{G}}_{0\,ab}(\omega)+\sum_{cd}\bm{\mathcal{G}}_{0\,ac}(\omega)\bm{\Sigma}_{cd}(\omega)\bm{\mathcal{G}}_{db}(\omega)\,.
	\label{eq:gorkov}
\end{equation}
where $\bm{\mathcal{G}}_{0}(\omega)$ is the known unperturbed one-body propagator associated with $| \Phi_0 \rangle$. Typically, $| \Phi_0 \rangle$, and thus $\bm{\mathcal{G}}_{0}(\omega)$, is obtained by solving the self-consistent mean-field Hartree-Fock-Bogoliubov (HFB) equations~\cite{Ring}. 

The one-body self-energy
\begin{equation}
	\bm{\Sigma}_{ab}(\omega)\equiv\begin{bmatrix}
	\Sigma_{ab}^{11}(\omega)&\Sigma_{ab}^{12}(\omega)\\
	\Sigma_{ab}^{21}(\omega)&\Sigma_{ab}^{22}(\omega)
	\end{bmatrix}
\end{equation}
encodes many-body correlations and is the key quantity to be eventually expanded and truncated at the working order of choice in actual calculations. The self-energy is expanded according to the so-called \textit{Algebraic Diagrammatic Construction} (ADC) expansion scheme developed in quantum chemistry~\cite{Schirmer83} and currently in use in nuclear physics. At present, Gorkov SCGF calculations are performed at the first non-trivial order, i.e., the ADC(2) level. While ADC(3) approximation provides state-of-the-art calculations in the Dyson SCGF framework~\cite{cipollone2013isotopic,raimondi2018} appropriate to doubly closed-shell nuclei, it remains to be formulated and implemented in Gorkov SCGF. Consequently, the self energy presently reads as~\cite{Soma:2011aj, Soma:2013ona}
\begin{subequations}
\begin{align}
	\Sigma^{11(2)}_{ab}(\omega)&=\hspace{-0.2cm}\sum_{k_1k_2k_3}\hspace{-0.2cm}\bigg\{\frac{\mathcal{C}_a^{k_1k_2k_3}(\mathcal{C}_b^{k_1k_2k_3})^*}{\omega-E_{k_1k_2k_3}+i\eta}+\frac{(\bar{\mathcal{D}}_a^{k_1k_2k_3})^*\bar{\mathcal{D}}_b^{k_1k_2k_3}}{\omega+E_{k_1k_2k_3}-i\eta}\bigg\}\,,\\
	\label{eq:firstcase}
	\Sigma^{12(2)}_{ab}(\omega)&=\hspace{-0.2cm}\sum_{k_1k_2k_3}\hspace{-0.2cm}\bigg\{\frac{\mathcal{C}_a^{k_1k_2k_3}(\mathcal{D}_b^{k_1k_2k_3})^*}{\omega-E_{k_1k_2k_3}+i\eta}+\frac{(\bar{\mathcal{D}}_a^{k_1k_2k_3})^*\bar{\mathcal{C}}_b^{k_1k_2k_3}}{\omega+E_{k_1k_2k_3}-i\eta}\bigg\}\,,\\
	\Sigma^{21(2)}_{ab}(\omega)&=\hspace{-0.2cm}\sum_{k_1k_2k_3}\hspace{-0.2cm}\bigg\{\frac{\mathcal{D}_a^{k_1k_2k_3}(\mathcal{C}_b^{k_1k_2k_3})^*}{\omega-E_{k_1k_2k_3}+i\eta}+\frac{(\bar{\mathcal{C}}_a^{k_1k_2k_3})^*\bar{\mathcal{D}}_b^{k_1k_2k_3}}{\omega+E_{k_1k_2k_3}-i\eta}\bigg\}\,,\\
	\Sigma^{22(2)}_{ab}(\omega)&=\hspace{-0.2cm}\sum_{k_1k_2k_3}\hspace{-0.2cm}\bigg\{\frac{\mathcal{D}_a^{k_1k_2k_3}(\mathcal{D}_b^{k_1k_2k_3})^*}{\omega-E_{k_1k_2k_3}+i\eta}+\frac{(\bar{\mathcal{C}}_a^{k_1k_2k_3})^*\bar{\mathcal{C}}_b^{k_1k_2k_3}}{\omega+E_{k_1k_2k_3}-i\eta}\bigg\}\,,
\end{align}%
\label{eq:SelfEn_2nd}
\end{subequations}
where the needed tensors are defined as
\begin{subequations}
\begin{align}
	\mathcal{C}_a^{k_1k_2k_3} &\equiv\frac{1}{\sqrt{6}}\big[\mathcal{M}_a^{k_1k_2k_3}-\mathcal{M}_a^{k_1k_3k_2}-\mathcal{M}_a^{k_3k_2k_1}\big]\,, \\
	\mathcal{D}_a^{k_1k_2k_3}
	&\equiv\frac{1}{\sqrt{6}}\big[\mathcal{N}_a^{k_1k_2k_3}-\mathcal{N}_a^{k_1k_3k_2}-\mathcal{N}_a^{k_3k_2k_1}\big]\,,
\end{align}%
\label{CandD}
\end{subequations}
where
\begin{subequations}
\label{MandN}
\begin{align}
	\mathcal{M}_a^{k_1k_2k_3}&\equiv \sum_{ijk} \bar{v}_{akij}\mathcal{U}_i^{k_1}\mathcal{U}_j^{k_2}\mathcal{V}_k^{k_3}\,, \\
	\mathcal{N}_a^{k_1k_2k_3}&\equiv \sum_{ijk} \bar{v}_{akij}\mathcal{V}_i^{k_1}\mathcal{V}_j^{k_2}\mathcal{U}_k^{k_3}\,,
\end{align}
\end{subequations}
as well as
\begin{equation}
	E_{k_1k_2\ldots k_n}\equiv\omega_{k_1}+\omega_{k_2}+\ldots+\omega_{k_n}\,.
\end{equation}

\subsection{Equation of motion}
\label{sec_equations}

Solving Eq.~\eqref{eq:gorkov} directly after inputting the working approximation to the self energy (Eq.~\eqref{eq:SelfEn_2nd}) is not a practical way to proceed. As a matter of fact, the Lehmann-like representation of the self-energy in Eq.~\eqref{eq:SelfEn_2nd} authorizes a rewriting of Eq.~\eqref{eq:gorkov} as an energy-independent matrix diagonalisation problem. The possibility to do so is at the heart of the ADC expansion schemes that preserves, at any ADC(n) order, the analytical properties of the exact self-energy encoded in its Lehmann-like representation~\cite{Schirmer83}. At the ADC(2) level, the rewriting processes as follows. Introducing two new tensors $\mathcal{W}$ and $\mathcal{Z}$ through
\begin{subequations}
\begin{align}
	\mathcal{W}_k^{k_1k_2k_3}&\equiv\sum_a\frac{(\mathcal{C}_a^{k_1k_2k_3})^*\mathcal{U}_a^k+(\mathcal{D}_a^{k_1k_2k_3})^*\mathcal{V}_a^k}{\omega_k-E_{k_1k_2k_3}}\,, \\
	\mathcal{Z}_k^{k_1k_2k_3}&\equiv\sum_a\frac{(\bar{\mathcal{D}}_a^{k_1k_2k_3})^*\mathcal{U}_a^k+(\bar{\mathcal{C}}_a^{k_1k_2k_3})^*\mathcal{V}_a^k}{\omega_k+E_{k_1k_2k_3}}\,,
\end{align}
\end{subequations}
Eq.~\eqref{eq:gorkov} is rewritten as
\begin{subequations}
\begin{align}
	\omega_k\mathcal{U}_a^k &=\sum_b\big[h_{ab}\mathcal{U}_b^k+\tilde{h}_{ab}\mathcal{V}_b^k\big]  \\
	&+\sum_{k_1k_2k_3}\big[\mathcal{C}_a^{k_1k_2k_3}\mathcal{W}_k^{k_1k_2k_3}+(\bar{\mathcal{D}}_a^{k_1k_2k_3})^*\mathcal{Z}^{k_1k_2k_3}\big]\,, \nonumber\\
	\omega_k\mathcal{V}_a^k &=\sum_b\big[\tilde{h}_{ab}^{\dagger}\mathcal{U}_b^k-h_{\bar{a}\bar{b}}^*\mathcal{V}_b^k\big] \\
	&+\sum_{k_1k_2k_3}\big[\mathcal{D}_a^{k_1k_2k_3}\mathcal{W}_k^{k_1k_2k_3}+(\bar{\mathcal{C}}_a^{k_1k_2k_3})^*\mathcal{Z}^{k_1k_2k_3}\big]\,, \nonumber
\end{align}
\end{subequations}
where $h$ and $\tilde{h}$ denote respectively the Hartree-Fock and Bogoliubov one-body fields~\cite{Soma:2011aj,Soma:2013ona}. The four above relations form a set of coupled equations for $\mathcal{U}$, $\mathcal{V}$, $\mathcal{W}$ and $\mathcal{Z}$ that can be recast under the form of a matrix eigenvalue problem
\begin{equation}
\Xi
\begin{bmatrix}
\mathcal{U} \\
\mathcal{V} \\
\mathcal{W} \\
\mathcal{Z}
\end{bmatrix}_k
=
\omega_k
\begin{bmatrix}
\mathcal{U} \\
\mathcal{V} \\
\mathcal{W} \\
\mathcal{Z}
\end{bmatrix}_k\,,
\end{equation}
where
\begin{equation}
\Xi \equiv
\begin{bmatrix}
h & \tilde{h} & \mathcal{C} & \overline{\mathcal{D}}^* \\
\tilde{h}^\dagger & -\overline{h}^* & \mathcal{D} & \overline{\mathcal{C}}^* \\
\mathcal{C}^\dagger & \mathcal{D}^\dagger & E & 0 \\
\overline{\mathcal{D}}^T & \overline{\mathcal{C}}^T & 0 & -E
\end{bmatrix}
\equiv
\begin{bmatrix}
\Xi^{(1)} & \Xi^{(2)} \\
\Xi^{(2)\dagger} & \mathcal{E}
\end{bmatrix}
\label{eq:scheme_matrix}
\end{equation}
is an energy-independent Hermitian matrix whose diagonalisation is equivalent to solving Gorkov's equation at the ADC(2) level whereas diagonalising $\Xi^{(1)}$ alone reduces to the HFB problem~\cite{Soma:2011aj}. Eigenvectors and eigenvalues of $\Xi$ provide the spectral amplitudes and poles of the one-body propagator, respectively. The solutions of the eigenvalue equation must be found iteratively and self-consistently while imposing a fixed average particle number~\cite{Soma:2013ona}.

\subsection{Computational cost}
\label{sec_IT_method}

As visible from Eq.~\eqref{eq:ham}, the input to a many-body calculation comes under the form of a set of mode-$n$ (i.e. $n$-index) tensors, i.e., $n=2$ for the kinetic energy $\{t_{pq}\}$, $n=4$ for the two-body interaction $\{\bar{v}_{pqrs}\}$, $n=6$ for the three-body interaction $\{\bar{w}_{pqrstu}\}$, etc. The storage cost of the Hamiltonian is thus dominated by the mode-$2k$ tensor defining the highest $k$-body interaction and scales as $N_b^{2k}$, where $N_b$ denotes the basis dimension of the one-body Hilbert space.

Given the input tensors, Eqs.~\eqref{CandD}-\eqref{eq:scheme_matrix} specify the set of intermediate tensors and tensor networks used to eventually access the unknown many-body tensor of interest, i.e., $\bm{\mathcal{G}}_{ab}(\omega)$. The fact that Gorkov's equation can be written as a matrix diagonalisation problem is favorable since it allows, in principle, to build an iterative process in a straightforward way. Starting from a Bogoliubov reference state and its spectroscopic amplitudes $\mathcal{U}_0$'s and $\mathcal{V}_0$'s, the matrix elements of $\Xi$ are computed. Diagonalising it, a new set of amplitudes and poles is obtained from which an updated version of the matrix elements of $\Xi$ are computed. Performing a second diagonalisation, the procedure can be repeated to activate the iterative process, which is finalized once a given convergence criterion has been satisfied.

\begin{figure}[t]
\centering
\includegraphics[width=1.0\columnwidth]{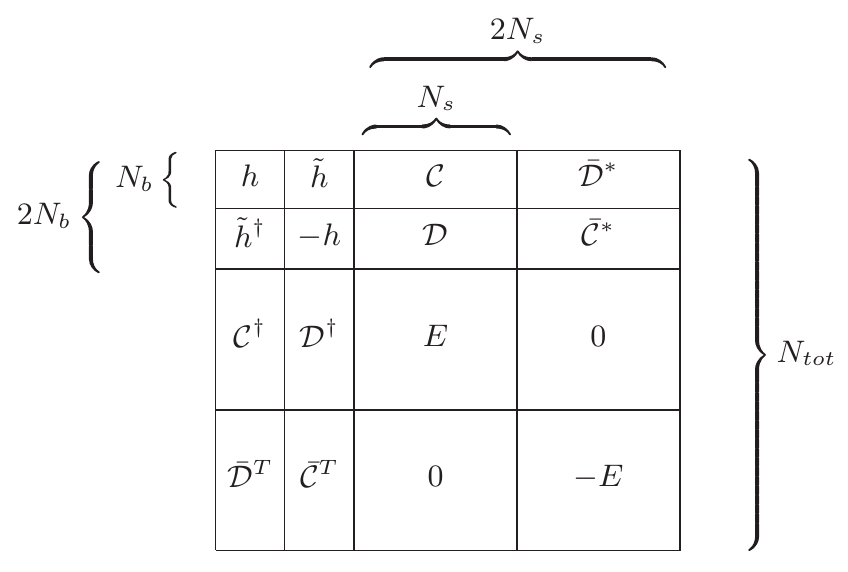}
\caption{Dimension scheme for the Gorkov matrix $\Xi$ (see text for details).}
\label{fig:GorkovMatrix}
\end{figure}

While the iteration process is straightforward in principle, a key difficulty relates to the fact that, given a fixed dimension $N_b$, the dimension $N_{\text{tot}}$ of the matrix to diagonalise is large and displays a combinatorial increase with the number of iterations in the self-consistent procedure. This feature is schematically illustrated in Fig. \ref{fig:GorkovMatrix}. While the HFB-like submatrix $\Xi^{(1)}$ is of fixed dimension $2N_b$, the rectangular (square) matrix $\Xi^{(2)}$ ($\mathcal{E}$) does not benefit from the same property given that it carries one (two\footnote{While $\mathcal{E}$ is diagonal at the ADC(2) level, it will become full at the ADC(3) level.}) three-pole index (indices) $k_1k_2k_3$. Indeed, while the number of poles is equal to $N_b$ when initializing the first iteration, it becomes equal to $N_{\text{tot}}=2(N_b+N_s)$ after the first diagonalisation of $\Xi$, where $N_s$ is essentially equal to $N^3_b$. Repeating the process with $N_{\text{tot}}$ initial poles unavoidably leads to a combinatorial increase of the number of poles and of the dimensionality of $\Xi$, making the problem intractable after a few iterations.

In an actual calculation, at each iteration, this explosion of the number of poles is controlled by two procedures.
The first one involves the use of an optimised reference state (OpRS)~\cite{Barbieri09}, i.e. an effective one-body propagator that is computed from the fully-correlated solution of Gorkov's equation but has the same number of poles as the HFB solution.
This allows to have the same $N_{\text{tot}}$ at every iteration while incorporating the effects of self-consistency.
The second one is a Krylov projection of the submatrix $\mathcal{E}$ prior to the diagonalisation of $\Xi$.
This allows to reduce Gorkov's matrix down to a size for which it can be exactly diagonalised via existing numerical routines.
Such a projection is performed by means of a dedicated Lanczos algorithm~\cite{Soma:2013ona}.

In spite of these procedures, the large dimensions of $\Xi$ (even at the first self-consistent iteration) generate a bottleneck when reaching out to doubly open-shell and/or heavier nuclei as well as when going to greater accuracy. While the former objective necessitates to increase the dimension of the one-body Hilbert space $N_b$, which mechanically increases $N_{\text{tot}}$, the latter objective leads in itself to a larger\footnote{Going from ADC(2) to ADC(3) does not actually increase the dimension of the matrix $\Xi$ but reduces tremendously the sparsity of the submatrix $\mathcal{E}$. This leads to a drastic increase of the non-zero matrix elements to compute and to the need for more involved techniques to perform the diagonalisation. Eventually going from ADC(3) to ADC(4) does indeed increase the size of $\Xi$ with additional sub-blocks running over five-pole indices $k_1k_2k_3k_4k_5$.} matrix $\Xi$ for a fixed value of $N_b$.

While the CPU-time needed to compute matrices $\mathcal{C}$ and $\mathcal{D}$, which can be significant, does not presently constitute a limiting factor, the RAM needed to store the matrix entries (possibly multiple times) does represent the true bottleneck. The logic to adapt IT techniques to GSCGF thus relies on the possibility to discard {\it a priori} the least relevant entries of $\Xi$ while maintaining an acceptable accuracy on the physical observables of interest.

\section{IT measures}
\label{sec_measures}

\subsection{Rationale}

\begin{figure}
\centering
\includegraphics[width=\columnwidth]{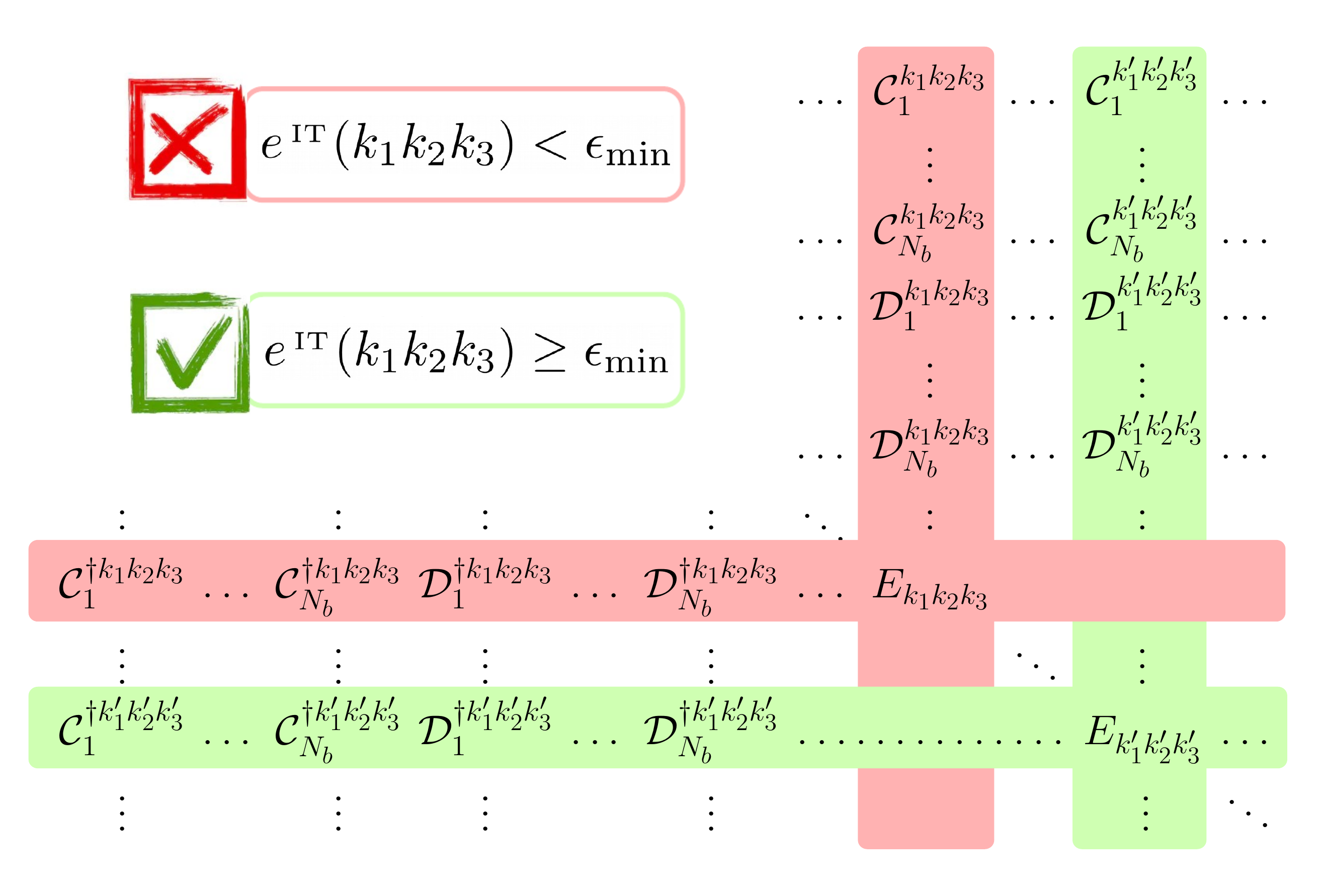}
\caption{Schematic illustration of the IT procedure applied to the construction of the matrix $\Xi$. The whole line and column associated with a given three-pole index are either kept or discarded.}
\label{fig:GorkovMatrix_IT}
\end{figure}

As schematically illustrated in Fig.~\ref{fig:GorkovMatrix_IT}, the goal is to discard {\it a priori} from matrix  $\Xi$ the least relevant three-pole configurations on the basis of a specific criterion. The criterion must combine, in one way or another, tensor entries $E_{k_1k_2k_3}$, $\{\mathcal{C}_a^{k_1k_2k_3}\}$ and $\{\mathcal{D}_a^{k_1k_2k_3}\}$ associated with  a given configuration $k_1k_2k_3$. The computation of $\{\mathcal{C}_a^{k_1k_2k_3}\}$ and $\{\mathcal{D}_a^{k_1k_2k_3}\}$ is an expensive task as it implies a sum over three single-particle indices (see Eqs. \eqref{CandD} and \eqref{MandN}). Still, the gain of IT techniques is that, in addition to reducing the dimension of the matrix to be diagonalised, the entries associated to discarded configurations do not have to be stored. 

The IT measure $e^{\,\,\!_{\text{IT}}}(k_1k_2k_3)$ is designed to gauge the relevance of each three-pole configuration on the basis of a less expensive method than the one of actual interest (SCGF in the present case). Given an appropriate IT measure, the configuration $k_1k_2k_3$ is kept only if 
\begin{equation}
e^{\,\,\!_{\text{IT}}}(k_1k_2k_3) \geq \epsilon_\text{min}\,,
\end{equation}
 where $\epsilon_\text{min}$ is a suitably chosen threshold. Conversely, it is discarded if
\begin{equation}
e^{\,\,\!_{\text{IT}}}(k_1k_2k_3) < \epsilon_\text{min} \,.
\end{equation}
The contribution of a discarded configuration to a given set of observables can eventually be re-added on the basis of the less costly method initially used to evaluate its importance. The whole process is performed on the fly without storing the corresponding matrix elements, which are finally erased.

The most naive example of IT measure is based on the hypothesis that the higher the unperturbed energy of a given elementary excitation, the less probable is the associated transition. By analogy with the HFB case, a first measure is thus introduced as
\begin{equation}
\ee{0}(k_1k_2k_3)\equiv\frac{1}{E_{k_1k_2k_3}}\,.  \label{0thorderIT}
\end{equation}
However, one would like to go beyond this first naive guess given that it does not include any information on the residual interaction encapsulated in tensors $\mathcal{C}$ and $\mathcal{D}$. Proposing, studying and testing optimal measures is the objective of the present work. 

Two categories of measures denoted as {\it method-driven} or {\it observable-driven} are presently considered. In the first case, the measure is directly inferred from the equations of motion at play in the method of interest (e.g. CI, CC, MBPT or SCGF in the present case), i.e., one is trying to evaluate the contribution of a given configuration to these equations of motion. As such, the IT measure does not focus on a particular observable. Contrarily, in the second case $e^{\,\,\!_{\text{IT}}}(k_1k_2k_3)$ estimates \textit{a priori} the contribution of a configuration to a particular observable. The associated IT approximation is expected to be optimal for the chosen observable whereas its impact on other observables has to be gauged down the road.

\subsection{Method-driven measures}

In the context of SCGF theory, method-driven measures naturally relate to the self-energy entering Dyson's or Gorkov's equation.
In the specific case of Gorkov SCGF formalism at the ADC(2) level, it is reasonable to assume that the most important configurations contributing to $\Sigma$ are those displaying the largest off-diagonal matrix elements (relative to their diagonal one) in the matrix $\Xi$. This expectation naturally leads to generalizing Eq.~\eqref{0thorderIT} by incorporating information on the residual interaction. 
Given that the normal self-energy is expected to dominate over the anomalous one, the analytical expression of $\Sigma^{11(2)}$ (Eq.~\eqref{eq:firstcase}) can be used as guidance to build the three following measures\footnote{In a spherically symmetric implementation, as the one considered here, one only needs to take into account $\mathcal{C}$ and $\mathcal{D}$ (not $\bar{\mathcal{C}}$ and $\bar{\mathcal{D}}$). This assumption is used throughout the present work. See Ref. \cite{Soma:2011aj} for details.}
\begin{subequations}
	\begin{align}
	e_{\,\,\!_{1A}}^{\,\,\!_{\text{IT}}}(k_1k_2k_3)&\equiv\hspace{-0.05cm}\sum_{ab}\bigg\{\frac{\mathcal{C}_a^{k_1k_2k_3}\mathcal{C}_b^{k_1k_2k_3}}{E_{k_1k_2k_3}}+\frac{\mathcal{D}_a^{k_1k_2k_3}\mathcal{D}_b^{k_1k_2k_3}}{E_{k_1k_2k_3}}\bigg\} \nonumber \\
	&= \hspace{-0.05cm}\frac{1}{E_{k_1k_2k_3}}\bigg\{\bigg|\sum_a\mathcal{C}_a^{k_1k_2k_3}\bigg|^2\hspace{-0.2cm}+\hspace{-0.05cm}\bigg|\sum_a\mathcal{D}_a^{k_1k_2k_3}\bigg|^2\bigg\}\,,
	\label{eq:mes1}\\	
	e_{\,\,\!_{1B}}^{\,\,\!_{\text{IT}}}(k_1k_2k_3)&\equiv\sum_{a}\bigg\{\frac{|\mathcal{C}_a^{k_1k_2k_3}|^2}{E_{k_1k_2k_3}}+\frac{|\mathcal{D}_a^{k_1k_2k_3}|^2}{E_{k_1k_2k_3}}\bigg\} \nonumber \\
	&= \frac{1}{E_{k_1k_2k_3}}\sum_a\Bigl\{|\mathcal{C}_a^{k_1k_2k_3}|^2+|\mathcal{D}_a^{k_1k_2k_3}|^2\Bigr\} \,,
	\label{eq:mes2}\\
	e_{\,\,\!_{1C}}^{\,\,\!_{\text{IT}}}(k_1k_2k_3)&\equiv\frac{1}{E_{k_1k_2k_3}}\Bigl\{|\mathcal{C}_1^{k_1k_2k_3}|^2+|\mathcal{D}_1^{k_1k_2k_3}|^2\Bigr\}\,.
	\label{eq:mes3}
	\end{align}
	\label{mes:first}
\end{subequations}
The first measure $e_{\,\,\!_{1A}}^{\,\,\!_{\text{IT}}}(k_1k_2k_3)$ directly derives from Eq.~\eqref{eq:firstcase} and involves sums of the components of the vector $v_a \equiv \mathcal{C}_a^{k_1k_2k_3}$. The second measure rather accounts for the norm of vector $v_a$. The third measure simplifies $e_{\,\,\!_{1B}}^{\,\,\!_{\text{IT}}}(k_1k_2k_3)$ by only retaining the (first) entry in the one-body basis\footnote{The calculations whose results are presented in Sec.~\ref{sec_results} are based on the use of the spherical harmonic oscillator basis. In this case, this choice corresponds to selecting the state with the lowest principal quantum number.}.

\subsection{Observable-driven measure}

As for a measure relating to a specific observable~\cite{Roth2009, roth2007ab, tichai2019pre}, the ground-state binding energy is presently considered. The formal task consists of expressing the observable of interest in terms of the entries in $\Xi$ associated with configuration $k_1k_2k_3$ that originate from the Lehmann-like representation of the self-energy. Furthermore, the IT measure must be based on an inexpensive method such that second-order perturbation theory is considered.
More specifically, its U(1)-breaking generalisation, i.e., Bogoliubov many-body perturbation theory (BMBPT)~\cite{Tichai:2018mll}, is employed.

Given the above two requirements, the ground-state binding energy must be expressed in perturbation theory in terms of the one-nucleon self energy. While the perturbative expansion of the ground-state correlation energy in terms of Goldstone diagrams is textbook material, it does not provide a functional dependence with respect to $\Sigma(\omega)$. This explicit connection is derived in \ref{edrivenmeasure} and constitutes an original formal development. Eventually, the energy-driven measure reads as
\begin{align}
e_{\,\,\!_{2}}^{\,\,\!_{\text{IT}}}(k_1k_2k_3)\equiv&\frac{1}{4}\sum_{k_4}\frac{\Big[\sum_a\big(\mathcal{V}_a^{k_4}\mathcal{C}_a^{k_1k_2k_3}+\mathcal{U}_a^{k_4}\mathcal{D}_a^{k_1k_2k_3}\big)\Big]^2}{E_{k_1k_2k_3k_4}}\,.
\label{eq:BMBPT_measure}
\end{align}

\section{Results}
\label{sec_results}

\subsection{Computational setting}

Gorkov SCGF calculations of closed- and open-shell nuclei are currently performed in $j$-scheme~\cite{Soma:2011aj}. The spherical harmonic oscillator basis is used to expand one-, two- and three-body operators. The basis truncation is set by the value of $e_{\text{max}} \equiv \text{max} (2n + \ell)$, where $n$ and $\ell$ are respectively the principal quantum number and the orbital angular momentum. Bases characterised by $e_{\text{max}}=7, 9, 11$ and 13, corresponding to 72, 110, 156 and 210 spherical shells, respectively, are employed in the present work. The two largest values typically allow to produce converged calculations of medium-mass nuclei with soft interactions. Three-body matrix elements are further truncated at $e_{3\text{max}}=16$ because of computational requirements, which again is a standard choice for medium-mass nuclei (see e.g. Ref.~\cite{Soma20} for a detailed analysis of basis convergence). Three-body forces are taken into account following the effective-interaction formalism of Ref.~\cite{Carbone:2013eqa} truncated at the two-body level~\cite{cipollone2015chiral}.

The \sat{} Hamiltonian~\cite{Ekstrom15} is employed unless stated otherwise.
\sat{} has been widely used in nuclear structure applications in recent years and thus represents a typical input for \textit{ab initio} expansion methods. In order to explore the possible dependence of the results on the softness and the type of Hamiltonian in use, three additional interactions evolved via the similarity renormalisation group (SRG) are considered in Sec.~\ref{sec:HD}: \sat{} evolved\footnote{In the case of \sat, SRG-evolution is known to induce substantial four-body (and possibly higher-body) forces that make the corresponding results not equivalent to the ones obtained with the original Hamiltonian. Being interested in the performance of IT relative to a given starting Hamiltonian (and not in generating fully physical results) this drawback does not constitute an issue for the present work.} down to $\lambda_{\text{SRG}} = 2.4 \text{ and } 2.0 \text{ fm}^{-1}$ and the recently introduced \lnl{} Hamiltonian~\cite{Soma20} evolved down to $\lambda_{\text{SRG}} = 2.0 \text{ fm}^{-1}$.
Present calculations make use of harmonic oscillator frequencies $\hbar \Omega = 20 \text{ and } 18$ MeV for \sat{} and \lnl{} respectively.

Finally, in the present work only the first self-consistent iteration is carried out (i.e., the so-called sc0 approximation~\cite{Soma:2013ona} is employed) with a fixed number $N_\text{L}=50$ of Lanczos iterations in the Krylov projection of matrix $\Xi$.
While the use of IT in a fully self-consistent (OpRS) calculation will have to be assessed in the future, we have explicitly checked that present results are independent of the particular choice of $N_\text{L}$~\cite{PorroMaster}.

\subsection{Characterisation of the results}
\label{sub:definitions}

To quantify the gain provided by the IT procedure, i.e., the resulting reduced dimension of matrix $\Xi$, the ratio $R$ is introduced as
\begin{equation}
R\equiv\frac{\#\text{ selected configurations}}{\#\text{ total configurations}}\,.
\end{equation}
Given that the size of $\Xi$ is driven by the size of $\Xi^{(2)}$, i.e., by the number of all possible three-pole configurations, $R$ essentially characterizes the relative number of triplets $k_1k_2k_3$ retained on the basis of the employed IT measure.

A criterion accounting for the quality of the IT result also needs to be defined. Following Ref.~\cite{tichai2019pre}, the observable-based figure of merit is proposed
\begin{equation}
\Delta O (R) \equiv\frac{|O(R)-O|}{|O|}\,,
\label{eq:delta}
\end{equation}
where $O\equiv O(1)$ stands for the observable computed without any truncation. If $R=1$, $\Delta O(1)=0$ and the full result is reproduced. Contrarily, $\Delta O(0)=1$ implies 100\% error. In the following, a 1\% error on the computed observable after IT ($\Delta O(R)=0.01$) is chosen as a reference for a good description. Such an error is typically of the order of (or smaller than) other sources of systematic uncertainty associated with the many-body truncation, the finite size of the model space or the construction of the input Hamiltonian~\cite{Soma21}.

\subsection{Assessment of IT measures}\label{sec:assessment}
\begin{figure}[!t]
\centering
\includegraphics[width=\columnwidth]{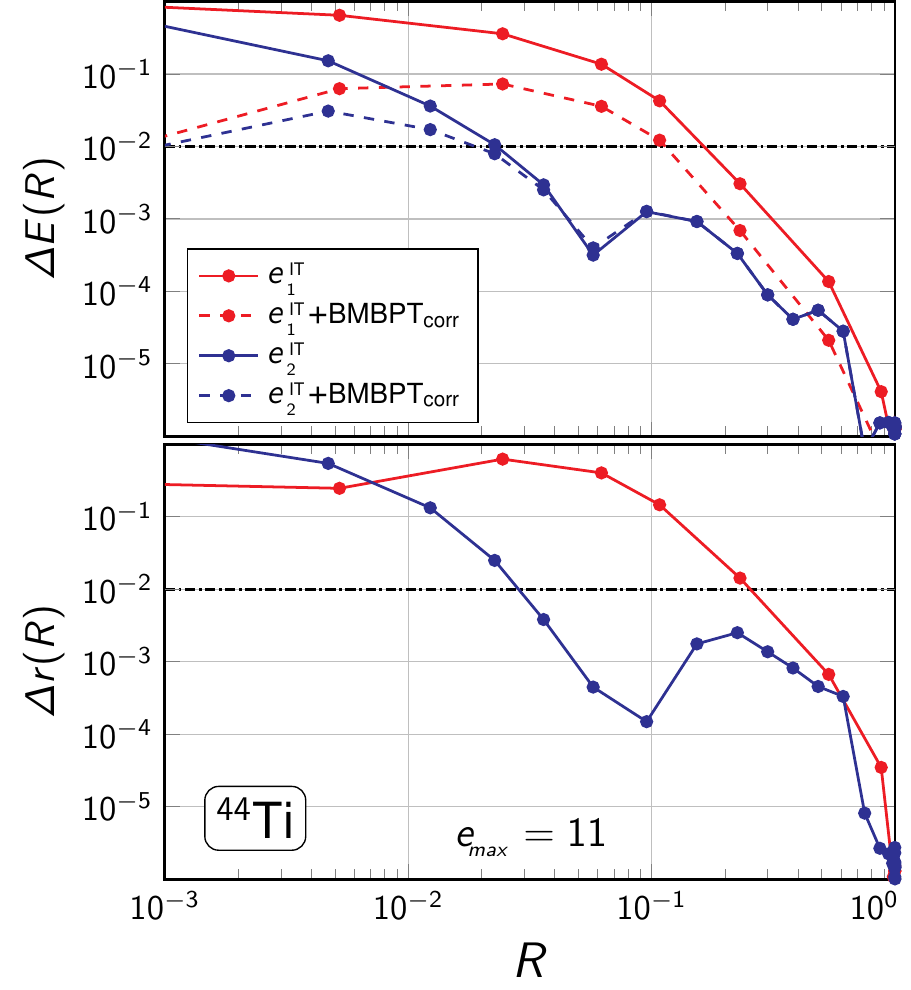}
\caption{Error of importance-truncated binding energy (top) and r.m.s. matter radius (bottom) as a function of the ratio $R$. Results for method- and energy-driven measures are shown as solid lines. 
For total energies, dashed lines correspond to self-consistent ADC(2) energies to which the contributions of the excluded configurations have been added perturbatively.  Calculations are performed for the doubly open-shell nucleus $^{44}$Ti in an $e_{\text{max}}=11$ basis.
}
\label{Fig_01}
\end{figure}
\begin{figure*}[!htbp]
\centering
\includegraphics[width=0.9\textwidth]{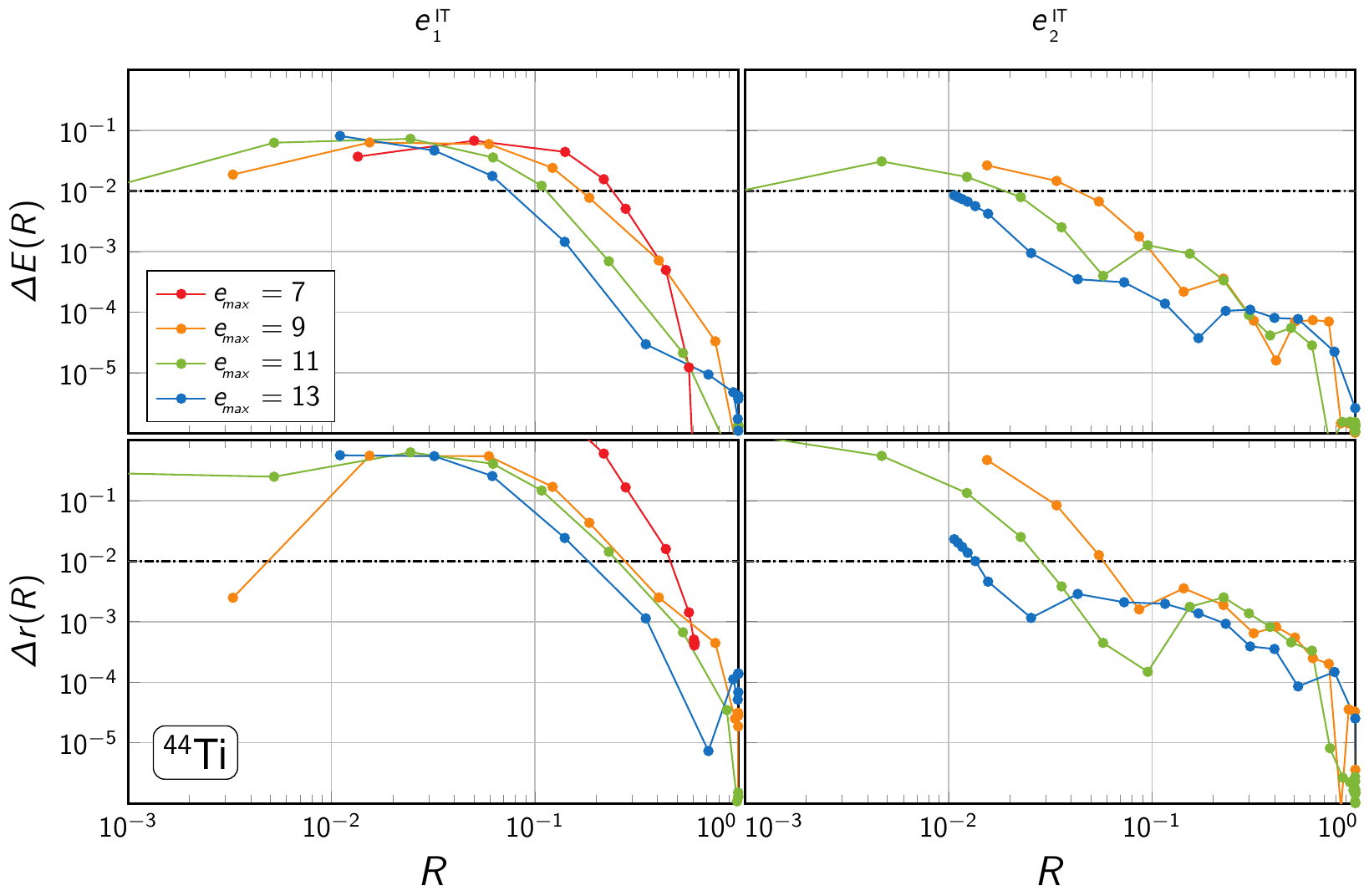}
\caption{Error of importance-truncated binding energy (top) and r.m.s. matter radius (bottom) as a function of the ratio $R$ for different values of the basis truncation $e_{\text{max}}$.
Results are show for the method-driven (left) and energy-driven (right) measures in the case of $^{44}$Ti.
}
\label{Fig_02}
\end{figure*}
In order to assess the general performance of the different measures introduced in Sec.~\ref{sec_measures}, the ground-state energy and root-mean-square (r.m.s.) matter radius of $^{16}$O is first computed~\cite{PorroMaster}. Although the naive $\ee{0}$ does eventually lead to a reduction of the number of poles, it proves very inefficient and is promptly discarded. The method-driven measures introduced in Eqs.~\eqref{mes:first} clearly improve the efficiency compared to $\ee{0}$. The three $\ee{1}$ measures yielding essentially indistinguishable results, the sole measure $\ee{1A}$, generically denoted as $\ee{1}$, is employed in the following\footnote{In principle, the measure $\ee{1C}$ is the cheapest to compute among the three. However, the cost of performing the sums over one one-particle index in Eqs.~\eqref{eq:mes2} and \eqref{eq:mes3} is negligible in practice such that the three $\ee{1}$ measures can be considered as equivalent also from the computational point of view.}.

Measures $\ee{1}$ and $\ee{2}$ are now extensively explored. Figure \ref{Fig_01} displays the figure of merit $\Delta O(R)$ (Eq.~\eqref{eq:delta}) for ground-state energies $E$ and r.m.s. matter radii $r$ in $^{44}$Ti. The two measures display two main differences. Measure $\ee{1}$ has a smooth and regular behaviour between the two limits $\Delta O \approx 0$ and $\Delta O \approx 1$ whereas $\ee{2}$ displays a less regular trend. At the same time, the gain achieved by the latter measure is about one order of magnitude higher for all considered values of $R$, thus providing a greater reduction of the dimension of $\Xi$. Consequently, the reduction factor allowed by $\ee{1}$ and $\ee{2}$ to meet the 1\% error (depicted as a dash-dotted line in the figures) on both the ground-state energy and charge radius is $R\approx 2 \cdot 10^{-1}$ and $R\approx 2 \cdot 10^{-2}$, respectively. While the optimal character of $\ee{2}$ for the energy could be anticipated, it is interesting to note that this measure is also very efficient for the calculation of another observable. This aspect will be further discussed in the following.

The expression defining $\ee{2}$, Eq.~\eqref{eq:BMBPT_measure}, essentially corresponds to the contribution of a given three-quasiparticle configuration ($k_1k_2k_3$) to the second-order perturbative correlation energy, see Eq.~\eqref{eq:BMBPT_en}. Therefore, by storing the computed $\ee{2}(k_1k_2k_3)$ values of the discarded three-quasiparticle poles, one automatically has access to the perturbative correction to the importance-truncated ADC(2) energy. 

Perturbatively corrected energies are plotted in Fig.~\ref{Fig_01} as dashed lines for both importance measures. One observes that the correction is significant for $\ee{1}$ throughout the whole range of $R$ values whereas it is only visible at small $R$ for $\ee{2}$. This reflects the fact that $\ee{2}$ precisely selects the configurations appearing in $\Xi$ according to their second-order MBPT energy contribution such that the correction to be added becomes smaller and smaller as $R$ increases. 

Once perturbatively corrected, both measures seem capable of recovering the $1\%$ accuracy for $R \approx 10^{-3}$, which would be one order of magnitude better than before. This gain is in fact not genuine. Indeed, for such tiny values of $R$, $\Xi$ essentially reduces to $\Xi^{1}$ such that the correlation energy is fully provided by the \textit{a posteriori}-added MBPT(2) correction, which itself happens to differ by about 1\% from the self-consistent ADC(2) value in the present calculation. In addition to the fact that this difference is somewhat accidental, other observables computed for $R \approx 10^{-3}$ are strongly degraded as exemplified\footnote{Other observables will be considered in Sec. \ref{sec:other}.} by the matter radius shown in the lower panel of Fig.~\ref{Fig_01}.
\begin{figure}[!h]
\centering
\includegraphics[width=0.91\columnwidth]{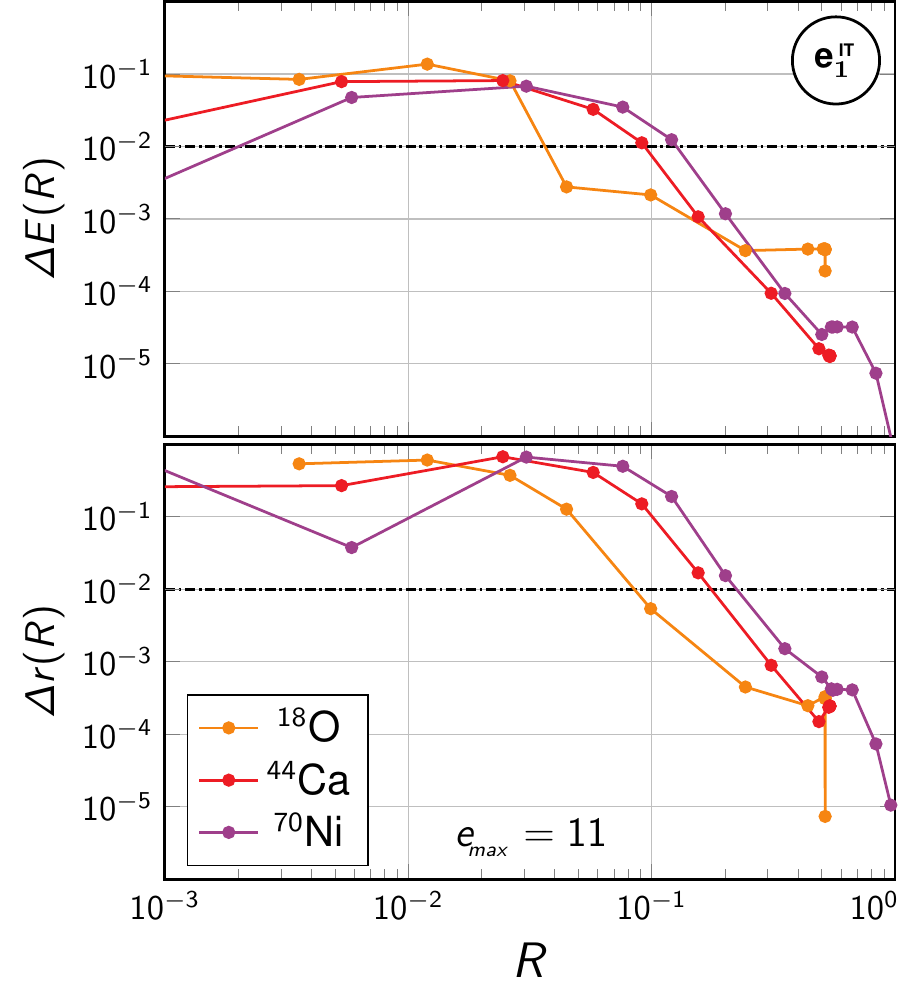}
\caption{Error of importance-truncated binding energy (top) and r.m.s. matter radius (bottom) as a function of the ratio $R$ for open-shell nuclei of different mass.
Results obtained with the method-drive measure $\ee{1}$ are displayed.}
\label{Fig_03}
\end{figure}
\begin{figure*}[!h]
\centering
\includegraphics[width=0.9\textwidth]{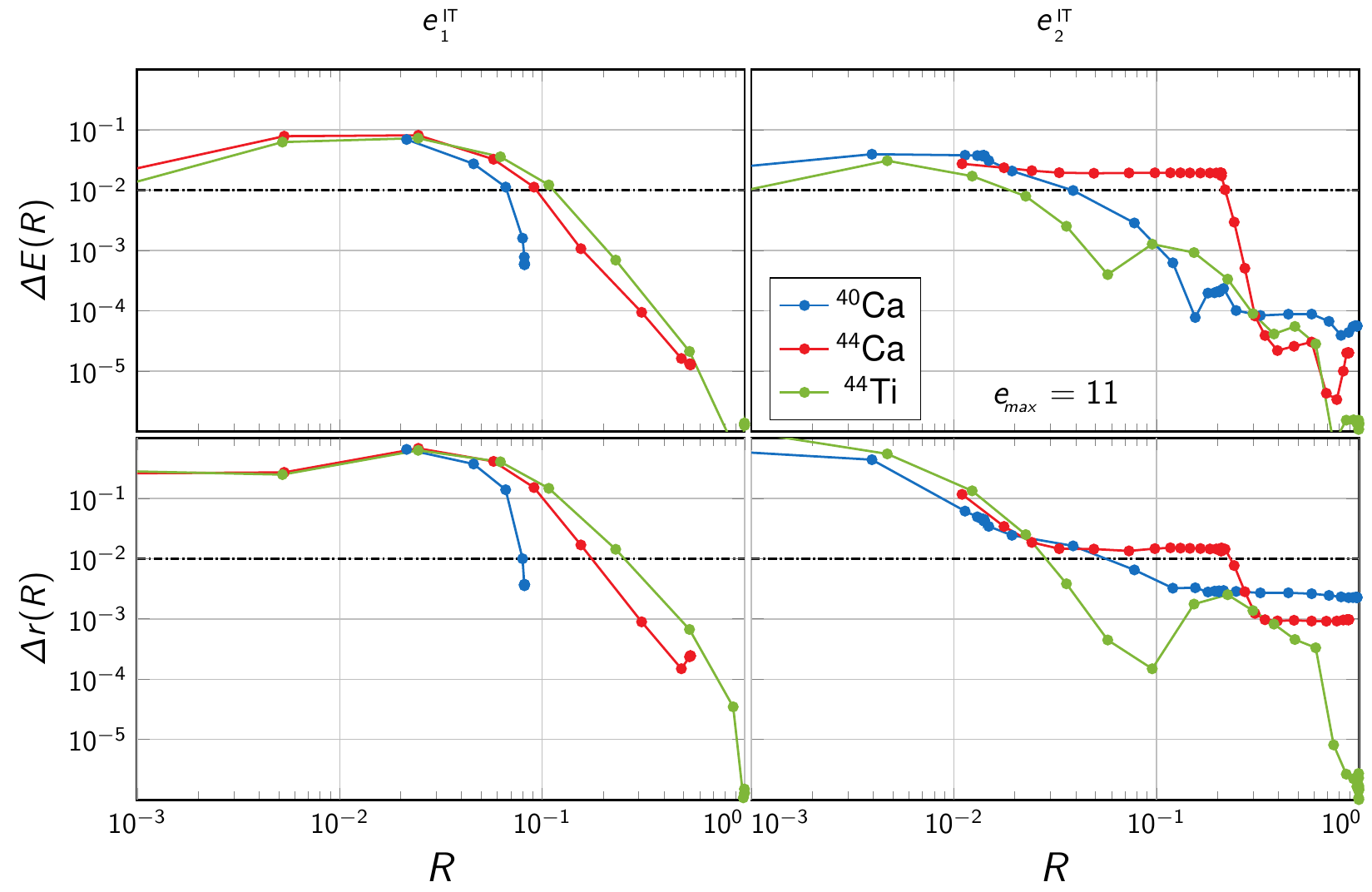}
\caption{Error of importance-truncated binding energy (top) and r.m.s. matter radius (bottom) as a function of the ratio $R$ for different closed- and open-shell nuclei.
Calculations performed with the method-driven (left) and energy-driven (right) measures
are shown.}
\label{Fig_04}
\end{figure*}

Eventually, genuine solutions delivering the targeted  $1\%$ accuracy are obtained for  $\ee{1}$ and $\ee{2}$ at $R\approx 10^{-1}$ and $R\approx 2.10^{-2}$, respectively. This is essentially identical to the results obtained without adding the perturbative correction. Still, unless specified otherwise, the perturbative correction to the energy is systematically incorporated in the remainder of the paper\footnote{In view of the above discussion, one must keep in mind that whenever the energy curve bends down as $R$ decreases, it signals that the perturbative correction dominates and that the one-body propagator solution of the IT Gorkov's equation tends towards the uncorrelated one.}.

\subsection{Dependence on model-space and mass}
\label{sec:modspace}

While results discussed in the previous subsection were obtained for $e_{\text{max}}=11$, they are reported on for $e_{\text{max}}$ varying from 7 to 13 in Fig.~\ref{Fig_02}. As expected~\cite{tichai2019pre}, the larger the size of the original tensors, the greater the gain, i.e., curves are systematically shifted to the left as $e_{\text{max}}$ increases. The trend is present for both observables and both measures, although not as clearly for $\ee{2}$ in the region $R \approx 1$ due its less smooth behaviour. 

As the dimension of ${\cal H}_1$ increases, the number of elementary, e.g.,  three-quasiparticle,  excitations incorporated grows combinatorially. Increasingly many of them, however, do not carry much or any importance in the effective description of the nucleus. In other words, while increasing $e_{\text{max}}$ is necessary to reach a better accuracy on the observables of interest, including all excitations thus allowed leads to processing a larger amount of irrelevant information.

The loss of efficiency carried by the $e_{\text{max}}$ truncation can be  quantified by looking at the upper panels of Fig.~\ref{Fig_02}. Regarding $\ee{1}$, the 1\% accuracy on the energy is reached at $R \approx 0.2$ for $e_{\text{max}}=7$ but at $R\approx 0.07$ for $e_{\text{max}}=13$. The gain is even larger for $\ee{2}$, which allows reaching the threshold at $R \approx 0.06$ ($R \approx 0.01$) in $e_{\text{max}}=9$ ($e_{\text{max}}=13$). The behaviour is qualitatively (for $\ee{1}$) or even quantitatively (for $\ee{2}$) similar for radii. In particular, results obtained with $\ee{2}$ reinforce the observation that the energy-driven measure performs surprisingly well for another observable.

Although the increased gain in large model spaces is expected to be a general effect for all nuclei, any quantitative consideration is necessarily system-dependent. 
To better put in perspective the results of Fig.~\ref{Fig_02} we analyse the variation associated to different mass regimes by performing calculations of the three singly open-shell nuclei $^{18}$O, $^{44}$Ca and $^{70}$Ni. Results for binding energies and matter radii relative to the $\ee{1}$ measure are presented in Fig. \ref{Fig_03}. It is easy to see that an intuitive phase-space effect linked to the different nuclear sizes is at play. For a fixed basis, lighter nuclei show a higher gain at a given ratio $R$ for both observables. Evidently, this relates to the fact that excitations with high values of $n$ and $\ell$ are less relevant for small mass numbers and thus can be massively discarded.

\subsection{Dependence on closed-/open-shell character}

Gorkov SCGF grasps static pairing correlations from the outset at the price of working with a reference propagator that has twice the number of poles compared to the one at play in Dyson SCGF theory.
Consequently, the number of allowed configurations is larger in Gorkov's formalism than in Dyson's one at any given truncation level\footnote{In Dyson SCGF, elementary excitations correspond to many-particle-many-hole configurations. For instance, at the ADC(2) truncation level, Dyson's counterpart to three-quasiparticle configurations are two-particle-one-hole and two-hole-one-particle combinations.}. While the larger number of configurations is essential for open-shell nuclei, Gorkov's basis is partly redundant in closed-shell systems where genuine pairing correlations disappear, i.e., a large number of three-quasiparticle poles become associated to a null amplitude. Given that these null (i.e. numerically tiny) amplitudes should be efficiently filtered out by IT techniques, and one expects the gain to be larger in doubly closed-shell systems than in open-shell ones when using a numerical code based on Gorkov's formalism.

\begin{figure}
\centering
\includegraphics[width=0.95\columnwidth]{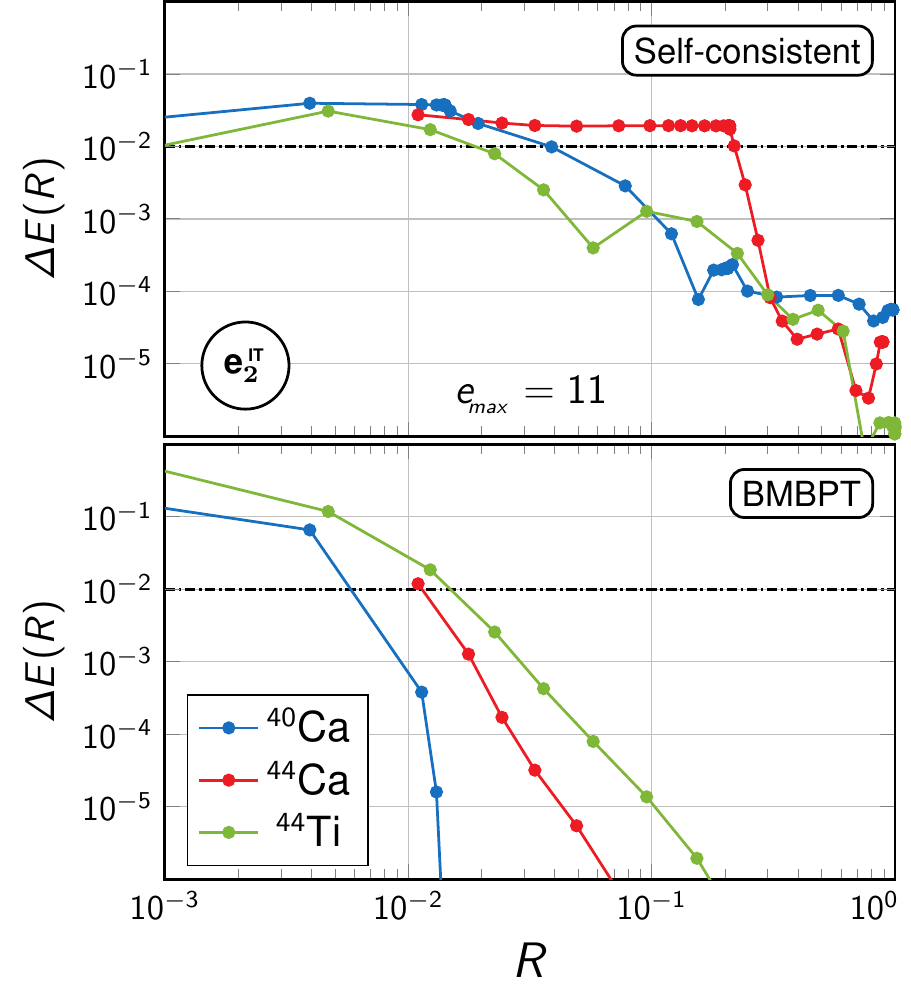}
\caption{Error of importance-truncated binding energy in the case of the energy-driven measure $\ee{2}$.
Self-consistent (top) and BMBPT (bottom) ground-state energies are displayed for $^{40}$Ca, $^{44}$Ca and $^{44}$Ti.}
\label{Fig_05}
\end{figure}
In order to confirm this expectation, results obtained in a doubly closed-shell nucleus, $^{40}$Ca, a singly open-shell nucleus, $^{44}$Ca, and a doubly open-shell nucleus, $^{44}$Ti, are compared in Fig.~\ref{Fig_05}. For $\ee{1}$, the gain is indeed more significant for $^{40}$Ca, i.e. the 1\% threshold is achieved with roughly twice less poles than in open-shell systems for both observables. In turn, $^{44}$Ca performs slightly better than $^{44}$Ti. As discussed above, this reflects the redundancy of Gorkov's formalism when applied to closed-shell systems. In physical terms, pairing correlations fragment the spectral strength already at the HFB level in open-shell nuclei. This gives rise to a higher density of states with non-zero spectroscopic amplitudes, in particular near the Fermi surface. In turn, this is encoded via a greater number of relevant configurations that cannot be neglected in the construction of the ADC(2) self-energy.

The picture changes when three-quasiparticle configurations are selected according to the energy-driven measure $\ee{2}$, as visible in right panels of Fig.~\ref{Fig_04}. Curves relative to $^{40}$Ca and $^{44}$Ti are somehow inverted, with the doubly open-shell system showing a more pronounced gain with respect to the doubly closed-shell one. Moreover, $^{44}$Ca displays a curious plateau around a few percent accuracy as the number of included poles is changed by more than one order of magnitude. This suggests that a peculiar feature arises when closed- and open-shell characters are mixed for protons and neutrons. The picture looks qualitatively the same for binding energies and radii.

To better understand the mechanism at play, Fig.~\ref{Fig_05} compares the IT results obtained with $\ee{2}$ for the self-consistent ADC(2) energy (already shown in Fig.~\ref{Fig_04}) and for its purely perturbative counterpart, i.e. the BMBPT(2) energy defined in Eq.~\eqref{eq:BMBPT_en}.
For BMBPT, $\Delta E(R)$ is very regular, does not display the plateau seen in $^{44}$Ca for the self-consistent calculation and restores the hierarchy between the three nuclei observed in Fig.~\ref{Fig_04} for the method-driven measure. Consequently, the reasons for the inversion between $^{40}$Ca and $^{44}$Ti as well as for the plateau observed in $^{44}$Ca are are to be found in the self-consistent character of the ADC(2) results.

\begin{figure}
\centering
\includegraphics[width=\columnwidth]{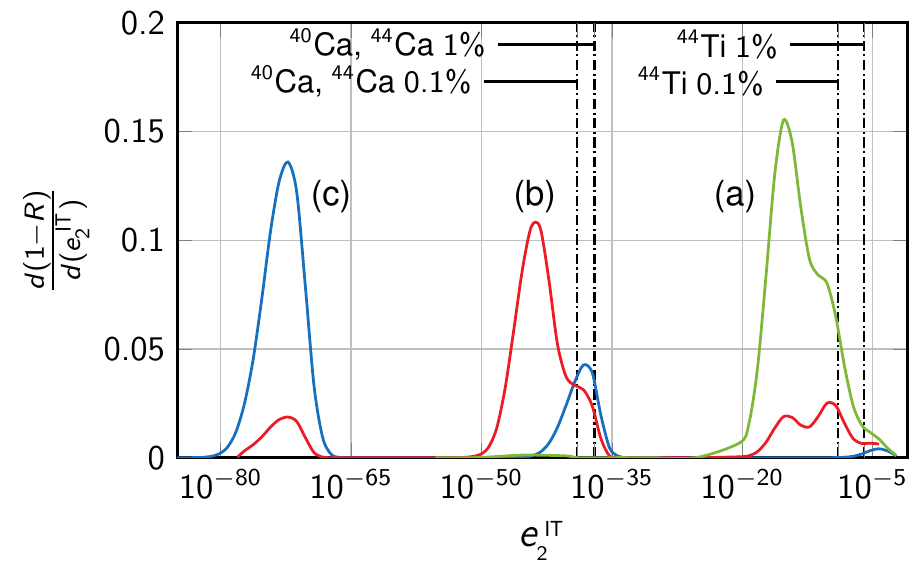}
\caption{Logarithmic distribution of three-quasiparticle configurations $(k_1k_2k_3)$ as a function of their $\ee{2}(k_1k_2k_3)$ value. 
Curves for $^{40}$Ca, $^{44}$Ca and $^{44}$Ti are displayed, with the colours as in Fig.~\ref{Fig_05}.
Critical values of $\ee{2}$ at which the self-consistent ADC(2) energy reaches a 1\% or a 0.1\% accuracy for the various systems are depicted as vertical dash-dotted lines.}
\label{Fig_06}
\end{figure}
\begin{figure*}[h]
\centering
\includegraphics[width=0.9\textwidth]{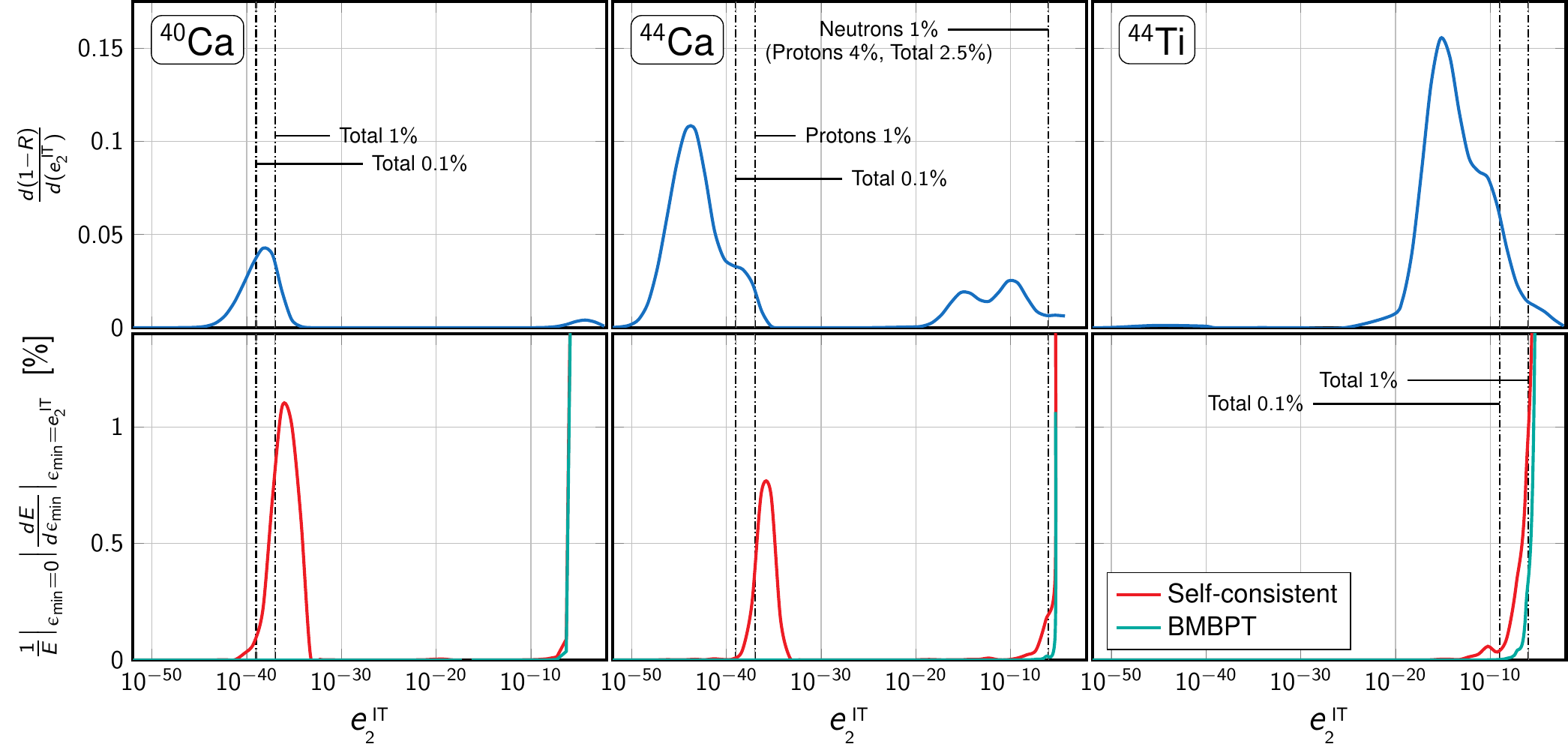}
\caption{Top: logarithmic distribution of three-quasiparticle configurations $(k_1k_2k_3)$ as a function of their $\ee{2}(k_1k_2k_3)$ value.
Bottom: differential contribution to the total binding energy (relative to the uncut result, in percent) as a function of the value $\epsilon_\text{min}$ at which the threshold for $\ee{2}$ is set. 
Curves for both BMBPT(2) and ADC(2) energies are shown. 
In all panels, critical values of $\ee{2}$ at which the self-consistent ADC(2) energy reaches a 1\% or a 0.1\% accuracy for the various systems are depicted as vertical dash-dotted lines.
For $^{44}$Ca, separate values for proton and neutron channels are also reported.}
\label{Fig_07}
\end{figure*}
Additional insight can be gained by analysing how three-quasiparticle configurations are distributed as a function of their contribution $\ee{2}$.
Distributions for $^{40}$Ca, $^{44}$Ca and $^{44}$Ti are shown in Fig.~\ref{Fig_06}.
Three distinct regions are visible
\begin{enumerate}
    \item[(a)] On the right, from 0 down to $\ee{2} \approx 10^{-20}$. Most of the strength of $^{44}$Ti is located here. $^{44}$Ca displays poles in the same energy range, although in much smaller number. $^{40}$Ca instead only shows a small peak in the right-hand side of the region.\\
    \item[(b)] In the centre, around $\ee{2} \approx 10^{-40}$. Here a large peak appears for $^{44}$Ca together with a smaller one for $^{40}$Ca. The doubly open-shell $^{44}$Ti only shows a shallow, residual concentration of strength.\\
    \item[(c)] On the left, around $\ee{2} \approx 10^{-70}$. Large part of the strength of $^{40}$Ca is found in this region. Additionally, a residual peak for $^{44}$Ca is observed.
\end{enumerate}
These different distributions can be qualitatively explained as follows.
After expanding the numerator in Eq.~\eqref{eq:BMBPT_measure} (and neglecting the sums over the single-particle indices\footnote{In the present $j$-coupled scheme, such sums eventually boil down to sums over principal quantum numbers only~\cite{Soma:2011aj}.}, both $a$ and the ones explicit in $\mathcal{C}$ and $\mathcal{D}$, as well as the cross product between the two terms) the energy-driven measure depends on the $\mathcal{U}$ and $\mathcal{V}$ amplitudes as
\begin{equation}
    \ee{2}(k_1k_2k_3) 
    \propto \sum_{k_4} [
    (\underbrace{\mathcal{U}^{k1} \mathcal{U}^{k2} \mathcal{V}^{k3}}_{\displaystyle \equiv \mathcal{A}^{k1k_2k_3}} \mathcal{V}^{k4})^2 +
    (\underbrace{\mathcal{V}^{k1} \mathcal{V}^{k2} \mathcal{U}^{k3}}_{\displaystyle \equiv \mathcal{B}^{k1k_2k_3}} \mathcal{U}^{k4})^2
    ] \, .
    \label{eq:peaks}
\end{equation}
In a closed-shell nucleus, at the HF level $\mathcal{U}^k$ and $\mathcal{V}^k$ assume values of 0 or 1, depending on the occupation of the state $k$, i.e. one has either $\{\mathcal{U}^k=0,\mathcal{V}^k=1\}$ or $\{\mathcal{U}^k=1,\mathcal{V}^k=0\}$.
It follows that $\mathcal{A}$ and $\mathcal{B}$ can be $1^3, 1^2 \cdot 0, 1 \cdot 0^2$ or $0^3$.
Given that in the sum in Eq.~\eqref{eq:peaks} there is always one $k_4$ for which $\mathcal{U}=1$ or $\mathcal{V}=1$, for each triplet $(k_1k_2k_3)$ the dominating term can have zero, two or four (in the worst case) powers of zero.
On the computer, a null HF occupation is not strictly zero but translates in a value of around $10^{-18}$.
Hence the three possibilities mentioned above, i.e., $\ee{2}(k_1k_2k_3) \propto 0^0, 0^2 \text{ or } 0^4$, give rise respectively to peaks\footnote{The position of the peak is driven by the powers of zero. The width is generated by the different values of interaction matrix elements, denominators and additional sums that have been neglected in this qualitative analysis.} (a), (b) and (c) in Fig.~\ref{Fig_06}.

In open-shell nuclei pairing correlations generate a fragmentation of the single-particle strength, i.e., fractional occupations, already at the HFB level.
This means that the $\mathcal{U}^k$ and $\mathcal{V}^k$ entering the second-order self-energy, hence the IT measures, are not strictly zero (i.e., $10^{-18}$ on the computer) but at least $\sim 10^{-5}$. 
For singly open-shell nuclei like $^{44}$Ca, this applies to one species only (neutrons in this case) and has the net effect of shifting some of the strength at higher values of $\ee{2}$. 
For doubly open-shell nuclei like $^{44}$Ti, there are hardly any null occupations at all and the vast majority of the distribution ends up being concentrated around the rightmost peak.

At this point one would be tempted to conclude that configurations belonging to peaks (b) and (c) are truly redundant, i.e., do not contribute to physical observables and that the only physically relevant region is the one on the right.
This is the case, by construction, as far as the BMBPT(2) energy is concerned.
Indeed, by looking solely at peak (a) in Fig.~\ref{Fig_06}, one deduces that many more configurations are needed to reach a given BMBPT(2) accuracy for $^{44}$Ti than for $^{40}$Ca, which is confirmed by the bottom panel of Fig.~\ref{Fig_05}.
However, by analysing the critical value of the $\ee{2}$ threshold $\epsilon_\text{min}$ at which the ADC(2) energy reaches a given accuracy (be 1\% or 0.1\%, depicted as vertical dashed lines in Fig.~\ref{Fig_06}), one finds that a non-negligible fraction of the poles belonging to peak (b) needs in fact to be included in doubly and semi-magic calcium isotopes.
This suggests that a number of seemingly (at the HFB level) irrelevant configurations must be necessarily incorporated in the self-consistent iterations in order to provide an accurate description of dynamical correlations in the system.

This finding is further confirmed by directly analysing the behaviour of the second-order energy as a function of the IT threshold, as done in Fig.~\ref{Fig_07}.
The top panels report the same curves as in Fig.~\ref{Fig_06}, with only the two rightmost peaks, i.e., peaks (a) and (b), being displayed. 
The bottom panels show instead the differential contributions to the second-order energy, i.e., the amount of added binding energy as one includes more and more configurations in matrix $\Xi$ according to the associated value of $\ee{2}(k_1k_2k_3)$ when going from right to left in the figure.
While BMBPT(2) energies receives sizeable contributions only for large values of $\ee{2}(k_1k_2k_3)$ in all three nuclei, self-consistent ADC(2) energies of $^{40}$Ca and $^{44}$Ca converge to an acceptable accuracy only after having included configurations located on the right-hand side of peak (b).
Because the residual energy generated by these configurations amount to a few percent of the total energy, including them is mandatory in the self-consistent calculation of closed-shell nuclei in order to reach our target accuracy.
For the singly open-shell $^{44}$Ca nucleus, in particular, reaching the central peak (b) requires to go first through the extended tail of peak (a) that carries a large number of configurations (see upper panel) contributing essentially nothing to the binding energy (see lower panel). This particular feature is responsible for the appearance of the plateau in Fig.~\ref{Fig_04}.

The above analysis makes clear that, while perturbation theory provides a proper tool to tailor an IT scheme, the behaviour of the observables computed within the non-perturbative method of interest does not necessarily follow the perturbative one. Consequently, the performance of the IT scheme is not only IT-measure dependent but is also many-body method dependent. Thus, the performance must be characterized thoroughly for each combination as is presently done for two IT measures within Gorkov SCGF theory.
\begin{figure}
\centering
\includegraphics[width=0.95\columnwidth]{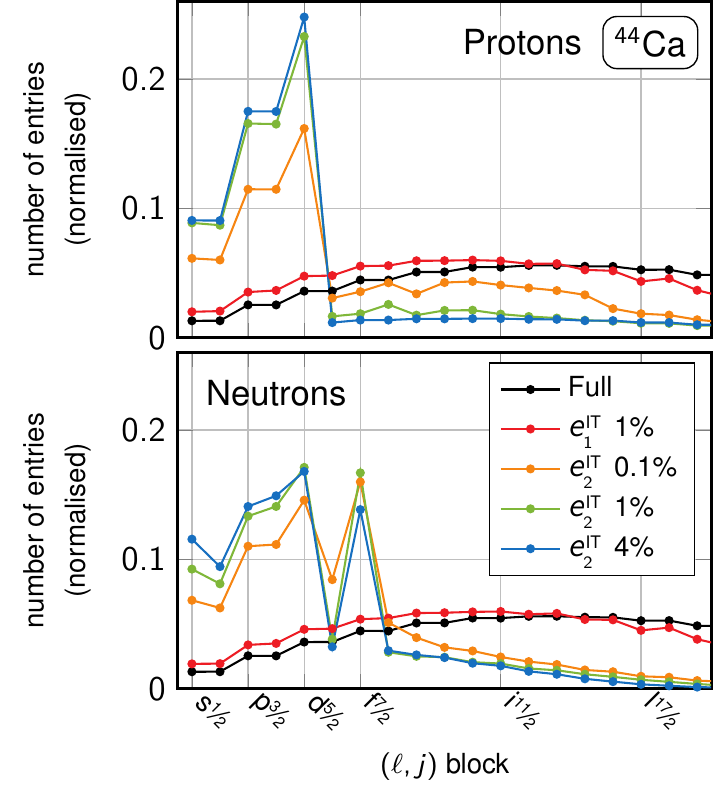}
\caption{Number of three-quasiparticle configurations in each of the ($\ell$,$j$) self-energy blocks after having applied different importance truncation measures.
All curves are normalised to one.
In the legend, the percentage refers to the corresponding accuracy of the ADC(2) energy.
Notice that for the energy-driven measure this percentage refers to separate proton and neutron contributions to $E$, i.e., $\ee{2}$ curves with the same colour in the two panels correspond to a different number of configurations in the proton and neutron Gorkov's matrices.
Not all x-axis labels are represented: only intermediate points are given for increasing values of $\ell$ and $j$.
All calculations were performed for the $^{44}$Ca nucleus in a model space with $e_{\text{max}}=11$.}
\label{Fig_10}
\end{figure}

To conclude, it is instructive to inspect the distribution of three-quasiparticle configurations as a function of the self-energy block\footnote{We recall that in the present $j$-scheme implementation Gorkov's matrix has a block-diagonal structure, with different blocks being labelled by $\ell, j$ and the isospin projection~\cite{Soma:2011aj}.} to which they contribute. 
These distributions are shown in Fig.~\ref{Fig_10} in the case of $^{44}$Ca for both full and importance-truncated calculations. The distribution in the full calculation simply reflects the available phase space that can be reached by three-quasiparticle excitations, constrained by the various conservation rules and the initial truncation of the one-body basis.

The method-driven measure cuts rather homogeneously across the different blocks, with lower angular momenta being slightly enhanced (i.e., less truncated) with respect to the higher ones.
This already points to the fact that the simple $e_\text{max}$ truncation at play in full, i.e., standard, calculations leaves room for improvement.
The picture, however, looks rather different if the energy-driven measure is used.
In this case the distribution strongly reflects the occupations in the various $(\ell, j)$ channels, with blocks corresponding to proton and neutron states below the Fermi surface being strongly enhanced with respect to the ones above.
Notice that the neutron $f_{7/2}$ channel, partially occupied already at the mean-field level, also contributes with a large number of three-quasiparticle configurations. 
These features revealed by the $\ee{2}$ distributions follow from the fact that this measure is intimately related to the one-body density matrix. 
Interestingly, when going below a given accuracy, say of 1\%, one observes a shift of strength from below to above the Fermi surface, with states that are unoccupied at the mean-field level acquiring a more significant weight. This feature is more marked for the non-superfluid protons and reveals the onset of significant dynamical correlations that fragment and redistribute the spectral strength. 
Consistently, the effect is weaker for superfluid neutrons, for which occupations around the Fermi surface are already smeared out by static pairing correlations.
By inspecting Fig.~\ref{Fig_07} again, one realises that the fragmentation in the proton channels is prompted by the inclusion of the configurations belonging to peak (b).
Eventually, as more and more poles are included, the $\ee{2}$ distributions in Fig.~\ref{Fig_10} will approach the full one and the occupation-driven structure will be washed out.
Finally, the picture emerging from the analysis of the energy-driven measure together with the fact that protons and neutron reach a given accuracy for rather different values of $R$ suggests that the use of a \textit{block-dependent} threshold could be explored in the future, in particular for semi-magic nuclei.

\subsection{Hamiltonian dependence}
\label{sec:HD}

Similarity renormalisation group techniques are routinely used to reduce the low-to-high momentum couplings in nuclear interactions, thus making many-body calculations more perturbative and more rapidly converging as a function of the basis size. 
In order to investigate how different SRG scales impact the IT performance two SRG-evolved versions of the \sat{} Hamiltonian are considered, namely with $\lambda_{\text{SRG}} = 2.4 \text{ and } 2.0 \text{ fm}^{-1}$, and compare the corresponding results with the ``bare'', i.e., SRG-unevolved, interaction used so far.
Furthermore, to test the possible dependence on the type of input Hamiltonian, the recently introduced \lnl Hamiltonian is additionally employed, also evolved down to $\lambda_{\text{SRG}} = 2.0 \text{ fm}^{-1}$.

\begin{figure}
\centering
\includegraphics[width=\columnwidth]{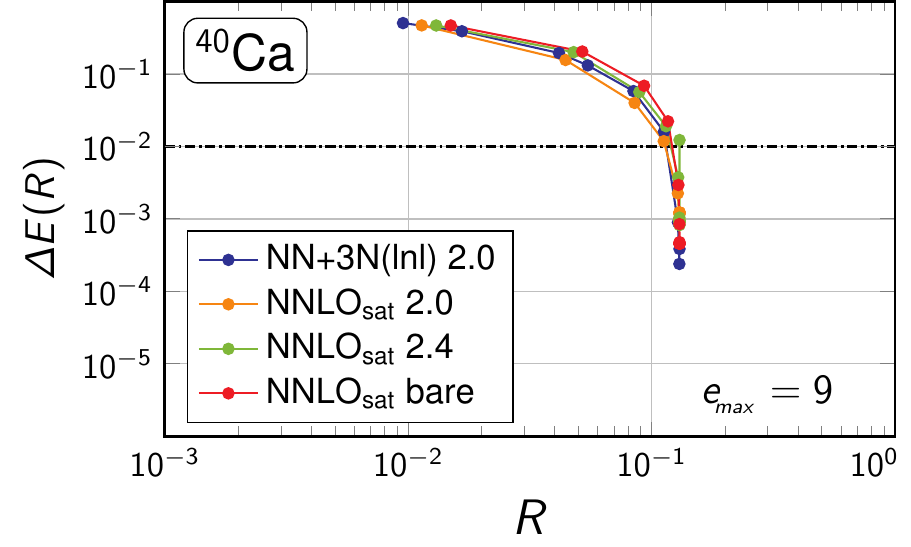}
\caption{Ground-state energy error as a function of $R$ for different input interactions using the method-driven measure $\ee{1}$. No BMBPT correction is included here.}
\label{Fig_11}
\end{figure}

Results focusing on the binding energy of $^{40}$Ca are displayed in Fig. \ref{Fig_11}.
The different curves follow the same trend and are almost superposed in a log-log scale. 
This shows that there are no major qualitative changes when softer interactions are employed, and points to a universal character of the efficiency of IT techniques.
This finding is not obvious \textit{a priori}, since different characteristic lengths $\lambda_{\text{SRG}}$ govern the details of how dynamical correlations are effectively built and, as a consequence, the relative weight of the different components of the objects encoding these correlations, e.g., the ADC(2) self-energy. For instance, while the second-order correlation energy $E^{(2)}$ grows as $\lambda_{\text{SRG}}$ is increased, the error $\Delta E(R)$ is not significantly affected.

\begin{figure}
\centering
\includegraphics[width=\columnwidth]{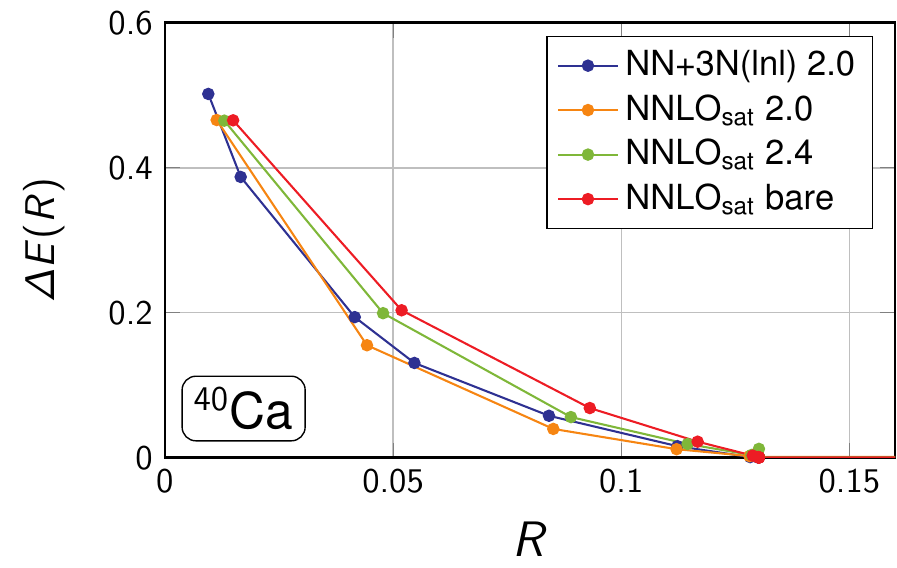}
\caption{Same as Fig.~\ref{Fig_11} but in linear scale.}
\label{Fig_12}
\end{figure}

A tangible effect is nevertheless present, as can be better appreciated in Fig.~\ref{Fig_12} where the same curves are displayed in a linear scale.
At a fixed ratio $R$, the importance-truncated energy is more accurate for smaller values of $\lambda_{\text{SRG}}$, consistently with the more perturbative character of the interaction.
This holds over a large range of $R$ values, with softer interactions displaying a more rapid decrease of the error as three-quasiparticle poles are added.
The two calculations characterised by different input Hamiltonians but with the same $\lambda_{\text{SRG}} = 2.0 \text{ fm}^{-1}$ show a very similar behaviour, suggesting that only the SRG scale (and not the interaction details) control the overall performance of IT.

Finally, we have also checked that these results are independent of the presence of three-body interactions and of their specific treatment. 
In particular, (i) all figures of merit show the same behaviour when only two-nucleon forces are taken into account and (ii) results for the extended Galitski-Migdal-Koltun sumrule~\cite{Carbone:2013eqa} are equivalent to the ones where the standard sumrule is considered (i.e. the expectation value of the three-body operator is neglected).

\subsection{Extrapolation procedure}\label{sec:extrapolation}

\begin{figure}
\centering
\includegraphics[width=\columnwidth]{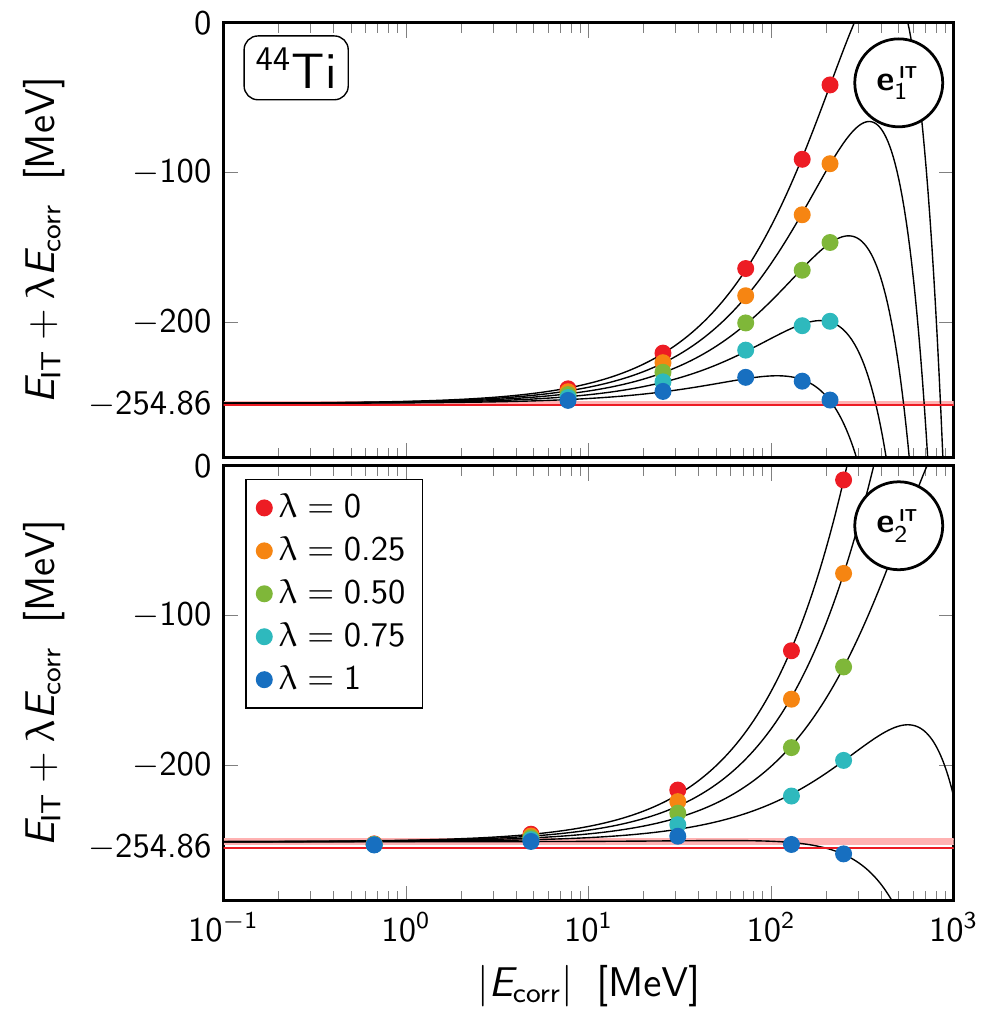}
\caption{Importance-truncated energy plus correction, Eq.~\eqref{eq:lambda_ext}, as a function of the perturbative correction $E_\text{corr}$. 
The top (bottom) panel shows results for the method-driven (energy-driven) measure.
In each panel, the pink band denotes the value (and corresponding error) of the extrapolated energy in the limit $E_\text{corr} \rightarrow 0$. The energy of the full calculations is represented by a red horizontal line.
Calculations are for $^{44}$Ti with $e_\text{max} = 11$.}
\label{Fig_fit}
\end{figure}

So far the IT procedure has been shown to deliver the full ADC(2) solution at a reduced cost, with an accuracy determined by the ratio of included over total three-pole configurations.
Eventually, the goal is to perform importance-truncated calculations in situations where the full computation is out of reach, i.e., where the reference solution is not at hand.
In such cases, how can one approach the full solution, and what can be concluded about the accuracy of the results?

The regular behaviour of the error functions presented in the previous sections suggests that some extrapolation procedure can possibly be devised, in particular for binding energies.
In Ref.~\cite{Roth2009} such a procedure was designed\footnote{In Ref.~\cite{Roth2009} the extrapolation technique was designed and employed within the frame of no-core shell model calculations as a function of $\epsilon_\text{min}$. Because of the self-consistent, highly non-linear, character of the ADC(2) equations, $\epsilon_\text{min}$ or $R$ do not provide suitable enough variables for the construction of a smooth function to be subsequently extrapolated. Instead,  $E_\lambda$ is presently studied as a function of the perturbative correction itself, i.e. the function $E_\lambda (E_\text{corr})$ is constructed and extrapolated to the limit $E_\text{corr} \rightarrow 0$. It can indeed be shown that $E_\text{corr}(\epsilon_\text{min})$ is a monotonous function going smoothly to zero as $\epsilon_\text{min} \rightarrow 0$~\cite{Buenker75}. It can thus play the role of a measure of the error itself, even in absence of the exact solution.} by constructing a family of energy sequences
\begin{equation}
    E_\lambda(\epsilon_\text{min}) \equiv E_{\,\,\!_{\text{IT}}}(\epsilon_\text{min}) + \lambda E_\text{corr}(\epsilon_\text{min})\, ,
\label{eq:lambda_ext}
\end{equation}
where $E_{\,\,\!_{\text{IT}}}(\epsilon_\text{min})$ is the IT energy at a given truncation threshold $\epsilon_\text{min}$ and $E_\text{corr}(\epsilon_\text{min})$ is the corresponding perturbative correction. Since $E_\text{corr}(\epsilon_\text{min}) \rightarrow 0$ whenever $\epsilon_\text{min} \rightarrow 0$, $E_\lambda(\epsilon_\text{min})$ recovers the full  energy $E$ in this limit independently of $\lambda$. Simultaneously fitting $E_\lambda(\epsilon_\text{min})$ for different values of $\lambda$ guarantees a stable and robust procedure to extrapolate the calculated points to the full energy $E$.

\begin{figure}
\centering
\includegraphics[width=0.9\columnwidth]{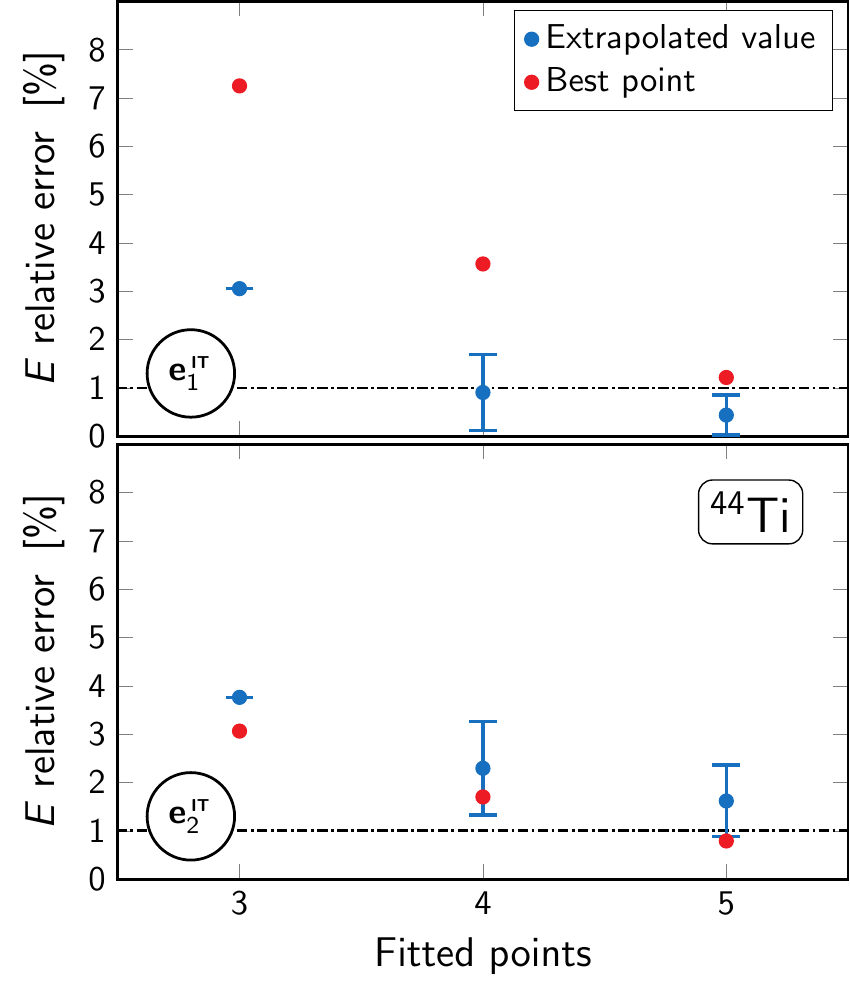}
\caption{Relative error on the ground-state energy extrapolated according to the procedure discussed in the text for different numbers of points used in the fit. The relative error for the best point included in the fit is also shown. The top (bottom) panel shows results for the method-driven (energy-driven) measure.
Calculations are for $^{44}$Ti with $e_\text{max} = 11$. The 1$\%$ accuracy threshold is indicated by an horizontal dash-dotted line.}
\label{Fig_fit_points}
\end{figure}

Figure~\ref{Fig_fit} displays the sequences of energies in $^{44}$Ti for five values of $\lambda$ between 0 and 1. 
Solid lines embody quadratic fits whereas the pink band delivers the corresponding extrapolated value $E_\text{extr}$ with associated error.
The red horizontal line denotes the full ADC(2) solution.
The two panels show results for the two measures $\ee{1}$ and $\ee{2}$.
In both cases the energy sequences smoothly approach the exact limit as $E_\text{corr} \rightarrow 0$.
A closer inspection however reveals that the extrapolation is more efficient for the method-driven measure $\ee{1}$. 
Indeed, for the latter one finds $E_\text{extr} = -253.7 \pm 1.0$ MeV while for $\ee{2}$ one has $E_\text{extr} = -250.7 \pm 1.9$ MeV.
The two are to be compared with the full energy $E = -254.8$ MeV.
The fact that $\ee{1}$ is more adapted to such an extrapolation procedure naturally reflects its smoother behaviour as a function of $R$ as visible in Figs.~\ref{Fig_02} and \ref{Fig_04}.

In Fig.~\ref{Fig_fit_points}, the error of the extrapolated value $E_\text{extr}$ (including uncertainty) is compared to the one of the best point used in the associated fit. The number of points employed in the fit is varied from three to five where one adds points from right to left on Fig.~\ref{Fig_fit} (energies quoted above correspond to the fit performed with five points). For both IT measures, the quality of the extrapolation improves when including more and more accurate points in the fit. One further observes that, while the extrapolation provides a gain compared to the best included point whenever $\ee{1}$ is employed, it is not the case for $\ee{2}$. This confirms that, while  $\ee{2}$ is a more efficient IT measure, the smoother error generated with $\ee{1}$ as a function of $R$ offers the possibility to extrapolate to the target 1$\%$ accuracy on the basis of points characterized by a lesser accuracy and that are thus cheaper to compute. This option will be optimized in future applications by computing more  points at "large" $E_\text{corr}$ to improve the fit that is currently performed with very few points. 

Last but not least, the extrapolation can in principle be applied to other observables, e.g., to r.m.s. radii considered previously. However, doing so requires to derive a perturbative estimate of the missing second-order contribution to r.m.s radii as a function of which the corresponding fit is to be performed.

\subsection{Other observables}\label{sec:other}

So far ground-state binding energies and r.m.s. matter radii have been investigated. It is useful to investigate the impact of the IT scheme on other quantities, namely density distributions and one-nucleon removal spectra. For this study, $\ee{2}$ is used at five representative truncation levels\footnote{Results obtained with the method-driven measure $\ee{1}$ yield analogous results.}.
Values corresponding to a $0.1\%$ and a $1\%$ accuracy on the ground-state energy are employed along with "$\ee{2}$ limit" corresponding to the leftmost point (i.e. $R \approx 10^{-3}$) in Fig.~\ref{Fig_04} for which the error reaches (accidentally) a $1\%$ accuracy. These three values are compared with the two extreme cases associated with the HFB and the full ADC(2) calculations.

\begin{figure}[b]
\centering
\includegraphics[width=\columnwidth]{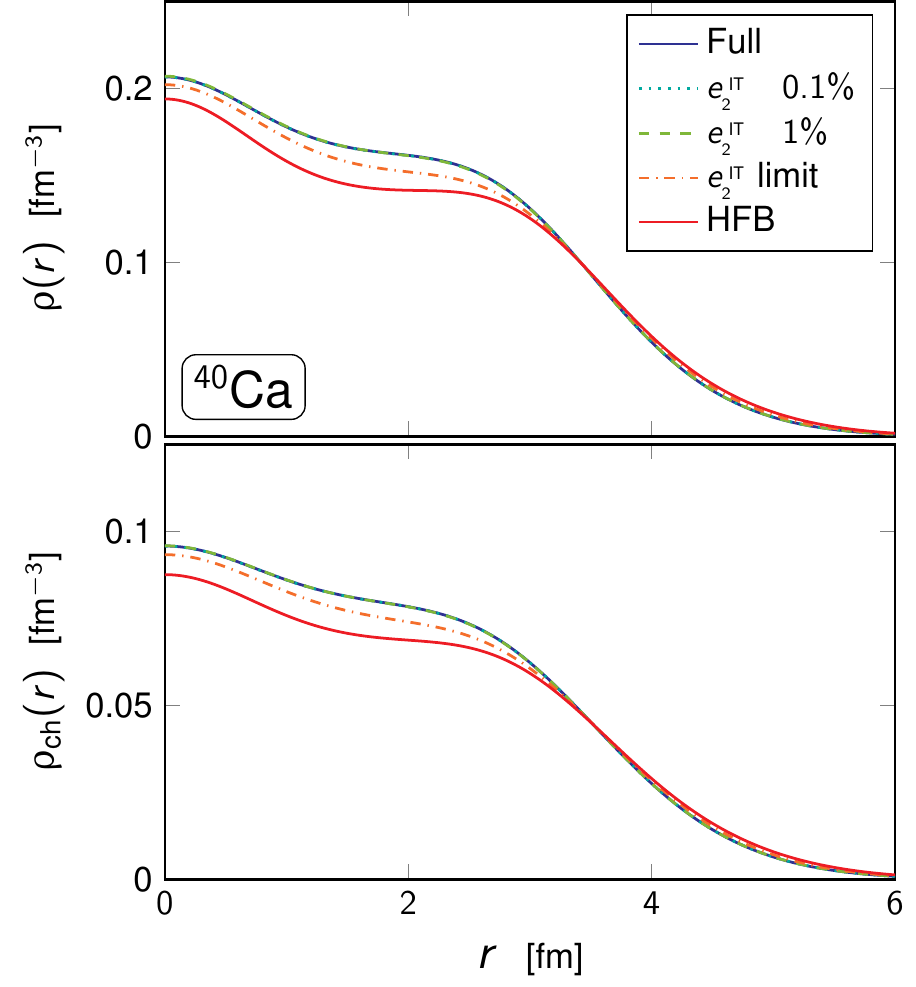}
\caption{Point-matter (top) and charge (bottom) density distributions of $^{40}$Ca for different $\ee{2}$ truncations (see text for details). Calculations were performed in an $e_{\text{max}}=11$ model space.}
\label{Fig_14}
\end{figure}
\begin{figure*}
\centering
\includegraphics[width=\textwidth]{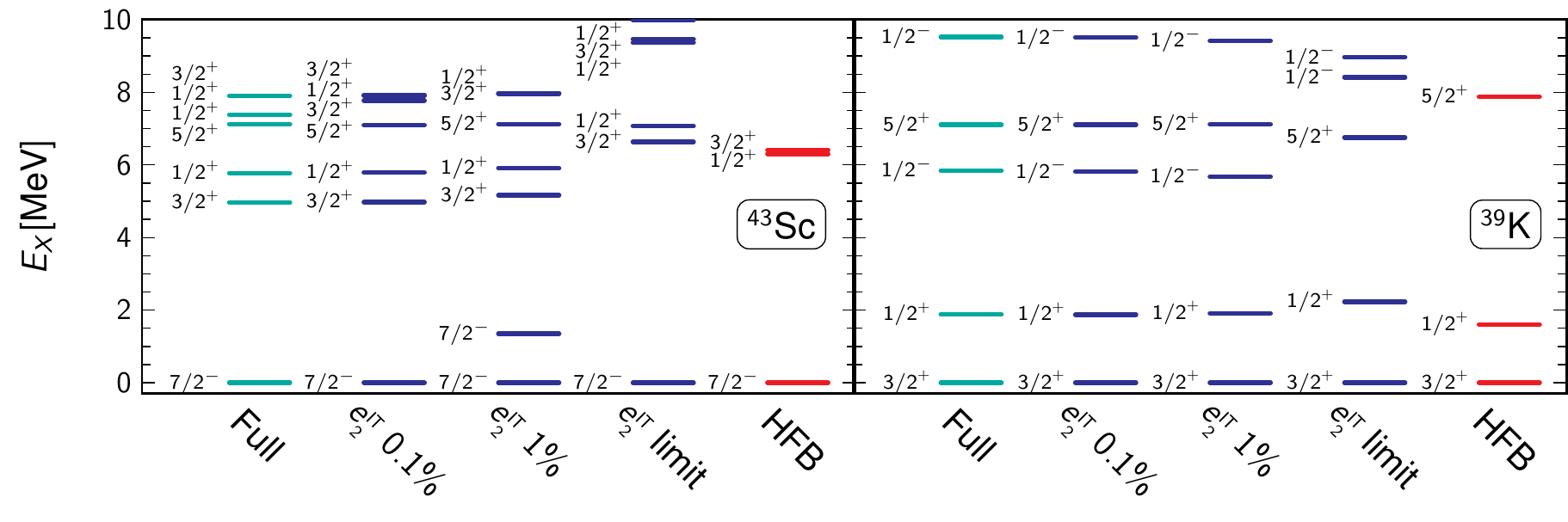}
\caption{One-proton removal spectra for $^{44}$Ti (left) and $^{40}$Ca (right) for different $\ee{2}$ truncations (see text for details).
In both panels states with $E_X<10$ MeV and spectroscopic factors $>1\%$ are shown.}
\label{Fig_15}
\end{figure*}
First, let us consider point-matter and charge densities.
In the SCGF formalism density distributions are computed starting from the density matrix, itself obtained from the one-body Green's function.
In order to calculate the charge density, corrections due to the finite size of the nucleons are further taken into account (see Ref.~\cite{Soma21} for details).
Figure \ref{Fig_14} shows the effect of the different $\epsilon_\text{min}$ thresholds on $\ee{2}$ on the two density profiles in the case of $^{40}$Ca. 
One observes that curves relative to $0.1\%$ and $1\%$ accuracy very well reproduce the full calculation in both cases, which is a direct confirmation that the full one-body propagator itself is very well reproduced.  Conversely, the "$\ee{2}$ limit" result falls in between the full and HFB curves, the latter yielding the poorest reproduction of the former as expected.

Second, one-proton removal energies from $^{44}$Ti and $^{40}$Ca, respectively corresponding to excitation spectra of $^{43}$Sc and $^{39}$K, are displayed in Fig. \ref{Fig_15}. Only states with spectroscopic factor larger than 1\% are shown. 
As for densities, one finds that whenever the total binding energy is reproduced at or better than the 1\% level the full one-nucleon removal spectra are well accounted for, i.e., for a 0.1\% (1\%) accuracy on the total energy, states carrying a large spectroscopic weight are shifted by at most few tens of keV (100-200 keV). Fragments with smaller spectroscopic factors might change more significantly, even in their relative ordering, especially for the 1\% accuracy.
This is the case, e.g., of the first excited $7/2^-$ state in $^{43}$Sc that acquires a larger spectroscopic factor in this particular calculation\footnote{One must keep in mind that such states are most probably not converged with respect to the many-body truncation when working at the ADC(2) approximation level.}. In contrast, the "$\ee{2}$ limit" spectrum presents significant differences with respect to the full one for all energy levels and rather tends towards the one of a HFB calculation. This further confirms that too drastic IT of the self-energy excessively impoverish the description of the dynamical correlations at play in nuclear systems. 

\subsection{Computational gain}
\label{sec_gain}

The ensemble of results presented in this study shows that by targeting a 1\% accuracy on the importance-truncated binding energy several other observables can be reproduced with a similar error. 
Such a value is smaller than the uncertainties typically affecting a state-of-the-art \textit{ab initio} calculation in this mass region (given a fixed input Hamiltonian)~\cite{Soma21} and can thus be assumed to be a safe and meaningful reference at which we can perform a quantitative assessment of the computational gain attainable via IT techniques. 

As discussed in Sec.~\ref{sec_IT_method}, the benefit of working with an importance-truncated Gorkov's matrix relates to both its RAM storage and the CPU time necessary to diagonalise it. 

\begin{figure}
\centering
\includegraphics[width=\columnwidth]{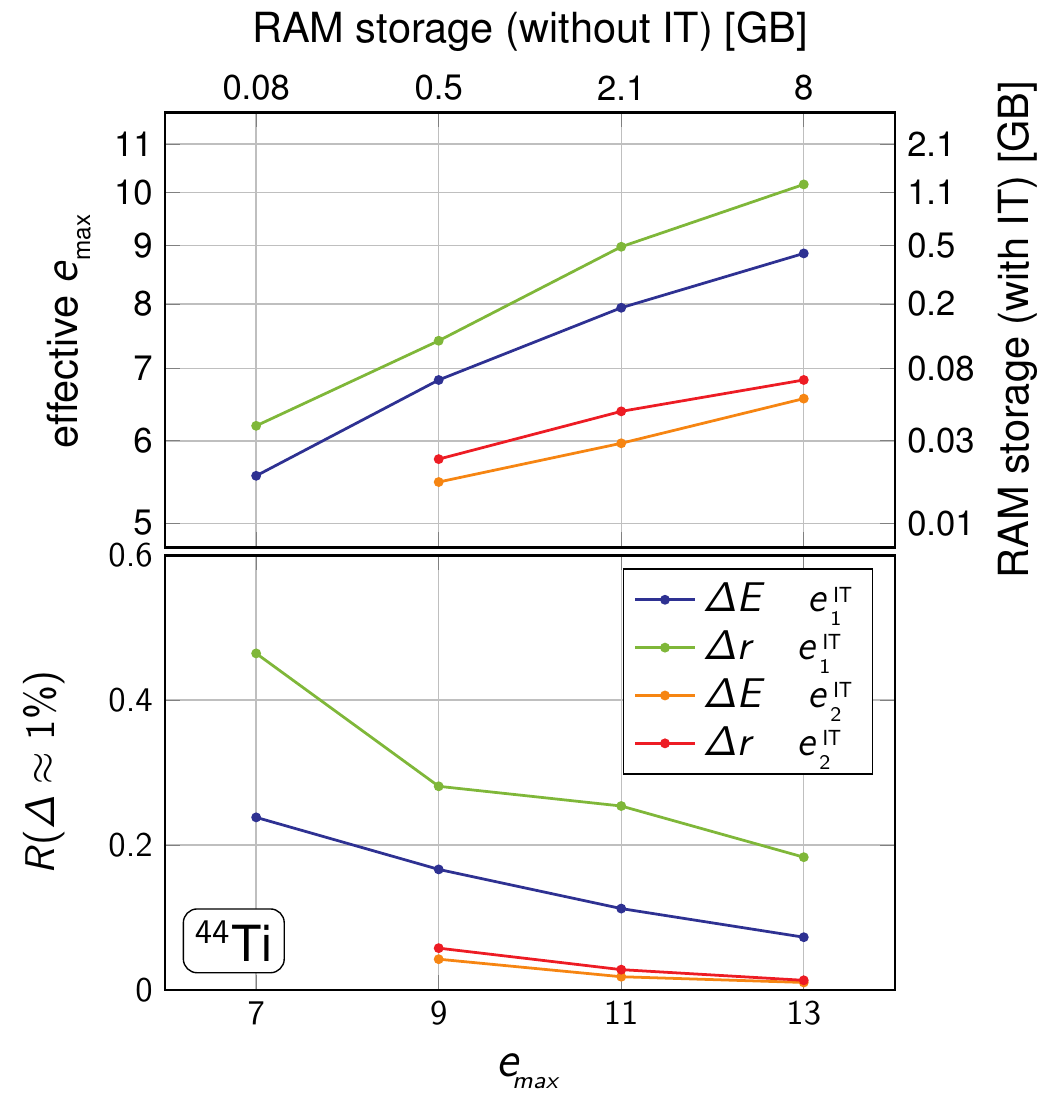}
\caption{Gain in memory storage allowed by IT as a function of the truncation of the basis of the one-body Hilbert space in $^{44}$Ti. 
Bottom: value of $R$ for which the error on a given observable crosses the 1\% threshold.
Top: requirements in RAM (right axis) and effective $e_\text{max}$ truncation (left, see text for details). For comparison, the original RAM requirements are displayed on the top axis.}
\label{Fig_16}
\end{figure}

The bottom panel of Fig.~\ref{Fig_16} summarises, as function of the one-body basis truncation $e_{\text{max}}$, the fraction of three-pole configurations that must be included to deliver a 1\% accuracy on total energies and matter radii.
Using the method-driven measure $\ee{1}$ already provides a significant reduction of the number of needed configurations, i.e., between 10 and 20\% of the total for realistic model spaces.
The energy-driven measure $\ee{2}$ allows a further reduction, with $R$ well below 5\% for both observables.
The corresponding gain in memory storage can be appreciated in the upper panel of the figure.
There, the original RAM storage of the full calculation for different $e_{\text{max}}$ is compared with the storage required after the application of IT, both in absolute terms (i.e. the total size of the memory allocation, right axis) and in terms of an \textit{effective} $e_{\text{max}}$ (i.e. the value of $e_{\text{max}}$ normally associated with this memory allocation, left axis).
As already remarked, the gain increases as one goes towards larger model-spaces. 
For instance, calculations in $e_{\text{max}}=13$ can be performed with memory resources lower than those required for an $e_{\text{max}}=7$ calculation when using the $\ee{2}$ measure.
This already constitutes an enormous gain and will be further enhanced when moving to even larger bases.

Such a benefit is certainly important but not (yet) critical at the current level of many-body truncation, i.e. an aggregate memory of 8 GB can be routinely handled in a $e_{\text{max}}=13$ full ADC(2) calculation.
However the situation will become drastically different at the ADC(3) level where calculations in even smaller model spaces will constitute a challenge. This is due to the fact that the off-diagonal elements of the submatrix $\mathcal{E}$ will become nonzero and will thus dramatically increases computational and storage requirements. Assuming a completely dense matrix\footnote{In reality the ADC(3) matrix has some sparsity, such that the present estimates can be considered as an upper limit.}, one will need as much as $\sim10^7$ GB of memory to store all elements of Gorkov's matrix when working in an $e_{\text{max}}=13$ basis, which is clearly out of reach of present computer facilities.
Applying IT techniques, one can reasonably hope to bring this memory allocation down to a few hundred GB, which remains challenging but could already be attained on state-of-the-art clusters.
Similar arguments apply each time one wishes to enlarge the size of the single-particle basis, e.g., to accommodate the further breaking of the SU(2) symmetry associated to angular momentum conservation or to include basis state that better describe the particle continuum like Bessel or Berggren states.
Memory-wise, the use of IT thus appears to be mandatory whenever a more sophisticated implementation of the self-energy is envisaged to meet future challenges, either in terms of many-body truncation or size of the one-body basis.

\begin{table}[]
	\centering
	\begin{tabular}{rccc}
		\hline
		\hline
		$\bm{e}_{\textbf{\text{max}}}$ & \textbf{Full} & $\bm{e_{\,\,\!_{1}}}^{\,\,\!_{\textbf{\text{IT}}}}$ & $\bm{e_{\,\,\!_{2}}}^{\,\,\!_{\textbf{\text{IT}}}}$ \\
		\hline
		\textbf{7} & 4 min & 1 min & - \\ 
		\textbf{9} & 37 min & 17 min & 14 min \\ 
		\textbf{11} & 4 h 3 min & 2 h 16 min & 2 h 5 min \\ 
		\textbf{13} & 29 h 36 min & 20 h 56 min & 17 h 14 min \\ 
		\hline
		\hline
	\end{tabular}
	\caption{CPU time needed for the computation and diagonalisation of the second-order self-energy matrix for $^{44}$Ti.
	For different model spaces, the full run is compared to importance-truncated calculations yielding a 1\% accuracy on the ground-state energy.}
	\label{tab:CPU}
\end{table}
When computing the importance measures, all elements of the ADC(2) matrix need to be evaluated. In this respect, eventually discarding some entries does not lead per se to a reduction of the CPU time when working at the ADC(2) level. Still, some gain comes from the fact that, afterwards, one deals with the diagonalisation of a smaller matrix.
In the present context, such diagonalisation makes use of a Krylov projection of the submatrix $\mathcal{E}$ as detailed in Ref.~\cite{Soma:2013ona}, which can become expensive when large model spaces are involved.
In Table~\ref{tab:CPU} the typical gain related to this computational step are exemplified in the case of $^{44}$Ti for both measures as a function of $e_\text{max}$.
Again, when going to ADC(3) such gain is expected to become much more significant.
Moreover, if the three-pole configurations can still be efficiently selected on the basis of the importance measures employed in the present work, an additional benefit will come from the fact that only the ADC(2)-like matrix elements of the ADC(3) matrix will have to be computed for the importance evaluation step. Correspondingly, the columns/lines of the ADC(3) matrix will only have to be evaluated in full for the tiny subset of configurations selected through the importance evaluation step. This feature would drastically impact the CPU time of Gorkov ADC(3) calculations.

\section{Conclusions}
\label{sec_conclusions}

\textit{Ab initio} methods have been established as reliable and predictive approaches to systematically attack the nuclear many-body problem. A robust link with underlying QCD is provided by the concomitant development of Hamiltonians within the context of chiral effective field theory~\cite{Hammer20} whereas both the development of innovative many-body methods and the growth in computational power have allowed to extend the reach of such calculations to mid-mass nuclei up to $A\sim100$ (or even beyond for some particular observables~\cite{Arthuis20}). However, associated numerical implementations quickly represent a formidable task due to the need to work in large Hilbert spaces~\cite{tichai2019pre}. In particular, they require to store and work with large and complex tensor networks, rapidly growing out of control even for state-of-the-art computer facilities. This feature is even more accentuated when dealing with non-perturbative methods~\cite{CompNucl2017}, requiring the iterative solution of a set of non-linear equations. 

In this context, importance truncation techniques aim at effectively reducing the CPU and storage costs of the numerical applications by discarding \textit{a priori} the less significant elements of the many-body tensors at play and, possibly, accounting perturbatively for their discarded contribution.
This \textit{a priori} estimate must be achieved via a less costly method than the one of interest, cutting down the computational cost of the problem and enabling to work in larger bases without paying the price of the non-perturvative computation in the full space. 

The present work addressed the development of IT techniques within the Gorkov SCGF formalism.
In this approach, the computational bottleneck preventing a systematic extension of the method to larger nuclei and/or greater accuracy lies in the storage and in the iterative diagonalisation of the second-order self-energy matrix (Sec. \ref{sec_equations}). 
In order to design and apply IT to GSCGF theory, a formal analysis was proposed in Sec. \ref{sec_measures} to properly relate the self-energy expansion of the dressed propagator to the perturbative series of the correlation energy. 
As an original result, a connection between the correlation energy and the Luttinger-Ward functional was established, which further led to the definition of new relations with the perturbative expansion. 
These theoretical efforts eventually allowed the definition of two different IT measures: a method-driven measure, accounting for the norm of many-body tensor entries, and an observable-driven measure, relying on the contribution of many-body tensor entries to the second-order correlation energy.

The two measures were tested in the calculation of several observables in closed- and open-shell systems. The energy-driven measure leads to an optimal gain, allowing to reduce down to $1\%$ the number of tensor entries while maintaining an accuracy of $1\%$ on total correlation energies, r.m.s. matter radii, as well as density distributions and one-nucleon separation energies.
For the same accuracy, the method-driven measure allows a lesser reduction of Gorkov's matrix but has the advantage of displaying a smoother dependence on the truncation parameter. 
The latter feature might turn out to be instrumental when the full solution is out of reach and one aims at extrapolating the IT results to the uncut limit.

Overall, both IT measures provide a robust tool to reduce computational and storage costs of non-perturbative many-body calculations. Although not (yet) critical at the current level of implementation of GSCGF theory, i.e., ADC(2), IT techniques will be pivotal in the next steps of development of the formalism, starting with the planned ADC(3) scheme. In particular, the ADC(3) approximation requires huge memory allocations that could hardly be handled without a systematic reduction of the size of Gorkov's matrix.

Alongside, the present study opens up several possible lines of future developments. 
A natural next step consists in deriving a perturbative expression for quantities other than the ground-state energy, e.g., spectral functions and radii.
This would enable one to add in perturbation the contributions from the discarded configurations also for these observables, as currently done for the correlation energy.
Finally, let us remark that many concepts and ideas introduced in the present work are not specific to the SCGF formalism or to its application to finite nuclei.
On the one hand, other non-perturbative many-body approaches like CC or IMSRG could benefit from similar developments.
On the other hand, these findings could be generalised to calculations of infinite nuclear matter or, e.g., to electronic systems.

\section*{Acknowledgements}

The authors wish to thank C. Barbieri and G. Col\`o for fruitful discussions, P. Arthuis for useful remarks on the manuscript and P. Navr\'atil for providing the interaction matrix elements used in the numerical calculations. A.P. is supported by the CEA NUMERICS program, which has received funding from the European Union's Horizon 2020 research and innovation program under the Marie Sk{\l}odowska-Curie grant agreement No 800945. 
A.T. was supported by the Deutsche Forschungsgemeinschaft (DFG, German Research Foundation) -- {Project-ID} 279384907 -- SFB 1245 and by the Max Planck Society.
Calculations were performed using HPC resources from GENCI-TGCC (Contracts No. A007057392, A009057392).

\appendix

\section{Energy-driven measure}
\label{edrivenmeasure}

\subsection{Interaction and correlation ground-state energies}
\label{def}

The objective is to define an energy-driven measure based on perturbation theory and expressed in terms of the tensors at play in ADC(2) Gorkov SCGF calculations. Let us thus introduce 
\begin{equation}
	\Omega(\lambda)\equiv \Omega_0+\lambda \Omega_1\,,
	\label{eq:ham_lambda}
\end{equation}
where $\lambda$ is the perturbation parameter leading the system from the uncorrelated state ($\lambda=0$) to the fully correlated (interacting) one ($\lambda=1$). The unperturbed grand potential $\Omega_0$ is taken to be of one-body character such that its ground-state is the Bogoliubov product state $\ket{\Phi_0}$, i.e.,
\begin{equation}
\Omega_0 \ket{\Phi_0}= \Omega^{0}_0\ket{\Phi_0}\,.
	\label{eq:unperturbed_shroedinger}
\end{equation}
Correspondingly, the ground state of $\Omega(\lambda)$ fulfils Schr\"odinger's equation 
\begin{equation}
	\Omega(\lambda)\ket{\Psi_0(\lambda)}=\Omega_0(\lambda)\ket{\Psi_0(\lambda)}\,.	\label{eq:perturbed_shroedinger}
\end{equation}
From Eqs.~\eqref{eq:unperturbed_shroedinger} and \eqref{eq:perturbed_shroedinger}, one obtains
\begin{align}
	\Omega_0(\lambda)&=\bra{\Psi_0(\lambda)}\Omega(\lambda)\ket{\Psi_0(\lambda)}  \nonumber \\
	&=\bra{\Psi_0(\lambda)}\Omega_0\ket{\Psi_0(\lambda)}+\lambda\bra{\Psi_0(\lambda)}\Omega_1\ket{\Psi_0(\lambda)}  \nonumber \\
	&\equiv \Omega_{\text{noint}} (\lambda)+ \Omega_{\text{int}}(\lambda)  \,.
	\label{eq:int_ener}
\end{align}
as well as 
\begin{align}
	\Omega_0(\lambda) &= \frac{\langle \Phi_0|\Omega|\Psi_0(\lambda)\rangle}{\langle \Phi_0|\Psi_0(\lambda)\rangle} \nonumber \\
	&=\frac{\langle \Phi_0|\Omega_0|\Psi_0(\lambda)\rangle}{\langle \Phi_0|\Psi_0(\lambda)\rangle} + \lambda \frac{\langle \Phi_0|\Omega_1|\Psi_0(\lambda)\rangle}{\langle \Phi_0|\Psi_0(\lambda)\rangle}  \nonumber \\
	&\equiv \Omega_0^0 + \Delta \Omega_0(\lambda)  \,.
	\label{eq:correl_ener}
\end{align}
The two above equations define the so-called interaction and correlation energies\footnote{The term ``energy'' is used everywhere even if, strictly speaking, the operator presently used is the grand potential and not the Hamiltonian.}, respectively, such that $\Omega_{\text{int}}(1)\equiv \Omega_{\text{int}}$ and $\Delta \Omega_0(1)\equiv \Delta \Omega_0$ for the fully interacting system. 

\subsection{Perturbative expansion of $\Delta \Omega_0$}
\label{exp}
 
The correlation energy is amenable to a perturbative expansion on the basis of Goldstone's theorem \cite{Brueckner1955, Goldstone1957}
\begin{equation}
	\Delta \Omega_0=\bra{\Phi_0}\Omega_1\sum_{n=0}^{\infty}\bigg(\frac{1}{\Omega_0^0-\Omega_0}H_1\bigg)^n\ket{\Phi_0}_{C}\,,
	\label{Goldstone}
\end{equation}
where the index $C$ denotes the fact that only so-called connected terms are retained. However, this expression does not explicitly relate the correlation energy to the propagator, the self energy and thus the many-body tensors at play in Gorkov SCGF calculations. 

\subsection{Linking $\Delta \Omega_0$ to the self energy}
\label{derivata}

To achieve the needed connection, one must first exploit Hellmann-Feynman's theorem~\cite{Feynman39} to relate the correlation energy to the interaction energy according to
\begin{equation}
	\Delta \Omega_0(\lambda)=\int_0^{\lambda}\frac{d\lambda'}{\lambda'}\Omega_{\text{int}}(\lambda')\,,
	\label{corr_ene_l}
\end{equation}
and further exploit the link the interaction energy entertains with the one-body propagator. 

Expressing $\Omega_{\text{int}}(\lambda)$ in terms of the two-body density matrix and exploiting the definition of the one-body self-energy connecting the one-body Green's function to the two-times reduction of the two-body Green's function\footnote{Using  $\Omega(\lambda)$, all quantities of interest (e.g. self-energy, one-body propagator, two-body density matrix...) become themselves $\lambda$ dependent.}, one easily obtains
\begin{equation}
\Omega_{\text{int}}(\lambda)= -\frac{1}{2} \text{Tr}\,\bigl\{\bm{\Sigma}(\lambda)\,\bm{\mathcal{G}}(\lambda)\bigr\}\,,
\label{e_int_l}
\end{equation}
where the triple trace over the one-body Hilbert space, Nambu space and the frequency defined through\footnote{The normalized trace accounts for a $1/2$ factor due to the $2\times2$ dimension of the quantities defined in Nambu space.}
\begin{equation}
\text{Tr} \,\equiv-\text{Tr}_{\mathcal{H}_1}\frac{1}{2}\text{Tr}_{\mathcal{G}}\oint\frac{d\omega}{2\pi i}\,,
\end{equation}
has been used to shorten the expression. Substituting Eq. \eqref{e_int_l} into Eq. \eqref{corr_ene_l} one obtains
\begin{equation}
	\Delta \Omega_0(\lambda)=-\frac{1}{2}\int_0^{\lambda}\frac{d\lambda'}{\lambda'}\text{Tr}\,\bigl\{\bm{\Sigma}(\lambda')\,\bm{\mathcal{G}}(\lambda')\bigr\}\,.
	\label{non_exp}
\end{equation}
Next, the self-energy is formally expanded in powers of the residual interaction according to\footnote{The quantity $\bm{\Sigma}_n(\lambda)$ indirectly depends on the coupling constant $\lambda$ given that it is itself a functional of the propagator $\bm{\mathcal{G}}(\lambda)$. Consequently,  $\bm{\Sigma}_n(\lambda)$ is \textit{not} the $n^\text{th}$-order contribution in a perturbative expansion, but simply the part of the exact functional $\bm{\Sigma}(\lambda)$ containing $(2n-1)$ propagators $\bm{\mathcal{G}}(\lambda)$ and $n$ interaction vertices~\cite{nozieres}.}
\begin{equation}
	\bm{\Sigma}(\lambda)=\sum_{n=1}^{\infty}\lambda^n\bm{\Sigma}_n(\lambda)\,,
	\label{Sigma_expanded}
\end{equation} 
such that Eq. \eqref{non_exp} becomes
\begin{equation}
	\Delta \Omega_0(\lambda)=-\frac{1}{2}\sum_{n=1}^{\infty}\int_0^{\lambda}d\lambda'\,\lambda'^{n-1}\text{Tr}\, \bm{\Sigma}_n(\lambda')\bm{\mathcal{G}}(\lambda')\,.
	\label{integral}
\end{equation}
Following standard steps~\cite{nozieres}, the integration over the coupling strength can be performed in Eq.~\eqref{integral} to eventually obtain\footnote{The ratio appearing in Eq.~\eqref{DE0_l} must be read as the matrix product $\bm{\mathcal{G}}(\omega,\lambda)\bm{\mathcal{G}}^{-1}_0(\omega)$, where $\bm{\mathcal{G}}^{-1}_0(\omega)$ is the inverse matrix of $\bm{\mathcal{G}}_0(\omega)$ in both the one-body Hilbert space ${\cal H}_1$ and the $2\times 2$ Nambu space.}
\begin{equation}
	\Delta \Omega_0(\lambda)=\Phi[\bm{\mathcal{G}}(\lambda)]+
	\text{Tr}\,\bigg\{\frac{\bm{\mathcal{G}}(\lambda)}{\bm{\mathcal{G}}_0}-1\bigg\}-\text{Tr}\,\bigg\{\ln\frac{\bm{\mathcal{G}}(\lambda)}{\bm{\mathcal{G}}_0}\bigg\}\,,
	\label{DE0_l}
\end{equation}
where the Luttinger-Ward functional~\cite{LuttingerWard1, LuttingerWard2} adapted to Gorkov's formalism was introduced
\begin{equation}
	\Phi[\bm{\mathcal{G}}(\lambda)]\equiv-\frac{1}{2}\sum_{n=1}^{\infty}\frac{\lambda^n}{n}\text{Tr}\,\bm{\Sigma}_n(\lambda)\bm{\mathcal{G}}(\lambda)\,.
	\label{eq:LW}
\end{equation}
Taking $\lambda = 1$ in Eq.~\eqref{DE0_l}, one obtains the fully interacting correlation energy $\Delta \Omega_0$ under the form of a functional of the fully interacting propagator $\bm{\mathcal{G}}(\lambda)$. The $\Phi$-functional $\Phi[\bm{\mathcal{G}}(\lambda)]$ invokes the convolution of $\bm{\Sigma}_n(\lambda)$, itself a functional of $\bm{\mathcal{G}}(\lambda)$, with $\bm{\mathcal{G}}(\lambda)$. As a result, $\Phi[\bm{\mathcal{G}}(\lambda)]$ is made out of the complete set of two-particle irreducible skeleton vacuum-to-vacuum diagrams whose lines denote the  fully interacting propagator $\bm{\mathcal{G}}(\lambda)$.

\subsection{Perturbative connection}
 \label{pert_connect}

In the present context, we are interested in a {\it perturbative} connection between the correlation energy and the one-body self energy. Starting from Eq.~\eqref{DE0_l}, it is easy to prove that $\Delta \Omega_0$ is stationary with respect to any variation of the propagator, i.e., $\bm{\mathcal{G}}(\omega)$ solution of Gorkov's equation delivers an extremum of $\Delta \Omega_0$. 

Concretely, this variational property translates into the possibility to alter the employed propagator without significantly affecting the correlation energy. This property was numerically validated in Ref.~\cite{dahlen2004}. Exploiting this feature, the correlation energy is presently {\it approximated} by evaluating its expression at the uncorrelated propagator rather at the interacting one, i.e., using
\begin{equation}
	\Delta \Omega_0\simeq\Delta \Omega_0[\bm{\mathcal{G}}_0]=\Phi[\bm{\mathcal{G}}_0]\,.
\label{LWG0}
\end{equation}
Based on this approximation, a perturbation-theory-like series is obtained for the correlation energy such that the $n^\text{th}$ order reads as
\begin{equation}
\Delta \Omega_0^{(n)}[\bm{\mathcal{G}}_0]=\Phi^{(n)}[\bm{\mathcal{G}}_0]=-\frac{1}{2n}\, \text{Tr}\,\bm{\Sigma}_n[\bm{\mathcal{G}}_0]\,\bm{\mathcal{G}}_0\,.
\label{eq:pert_Green}
\end{equation}
One must note that Eqs.~\eqref{Goldstone} and~\eqref{LWG0} do not provide, in general, the same expansion, i.e., while the first one is exact when summed to all orders, the second one is approximate. However, employing an uncorrelated propagator $\bm{\mathcal{G}}_0$ associated with the solution of the HFB mean-field equations, one can prove that the two expansions match exactly up to second-order. This exact matching has been validated numerically and is used in the following to compute the second-order correlation energy within Gorkov SCGF formalism.

The last step of the derivation thus consists of exploiting Eq.~\eqref{eq:pert_Green} for $n=2$ and express the result in terms of the many-body tensors making up the ADC(2) self energy. For a spherically symmetric implementation, the following identities hold
\begin{subequations}
\begin{align}
	\text{Tr}\{\Sigma^{12}\mathcal{G}^{21}\}&=\text{Tr}\{\Sigma^{21}\mathcal{G}^{12}\}\,,\\
	\text{Tr}\{\Sigma^{11}\mathcal{G}^{11}\}&=\text{Tr}\{\Sigma^{22}\mathcal{G}^{22}\}\,,
\end{align}
\end{subequations}
eventually leading to the second-order correlation energy expression
\begin{align}
\Delta \Omega_0^{(2)}[\bm{\mathcal{G}}_0]=&\frac{1}{4}\sum_{ab}\oint\frac{d\omega}{2\pi 1}\bigl\{\Sigma^{11(2)}_{ab}(\omega)\mathcal{G}_{0\,ba}^{11}(\omega) \nonumber \\
& \hspace{1.5cm} +\Sigma^{12(2)}_{ab}(\omega)\mathcal{G}_{0\,ba}^{21}(\omega)\bigr\}
\label{eq:e_2_gor}\,.
\end{align}
Inserting Eqs.~\eqref{Gorkov_prog_Lehmann} and \eqref{eq:SelfEn_2nd} into Eq.~\eqref{eq:e_2_gor}, the integration over the frequency can be performed to generate the needed form
\begin{align}
\Delta \Omega_0^{(2)}[\bm{\mathcal{G}}_0]&=\frac{1}{4}\sum_{\substack{k_1k_2k_3k_4\\ab}}\bigg\{\frac{{\mathcal{V}}_b^{k_4}{\mathcal{V}}_a^{k_4}\mathcal{C}_a^{k_1k_2k_3}\mathcal{C}_b^{k_1k_2k_3}}{E_{k_1k_2k_3}+\omega_{k_4}} \nonumber \\ 
&\hspace{1.2cm} +\frac{\mathcal{U}_b^{k_4}\mathcal{U}_a^{k_4}{\mathcal{D}}_a^{k_1k_2k_3}{\mathcal{D}}_b^{k_1k_2k_3}}{E_{k_1k_2k_3}+\omega_{k_4}} \nonumber \\ 
&\hspace{1.2cm} +\frac{{\mathcal{U}}_b^{k_4}{\mathcal{V}}_a^{k_4}\mathcal{C}_a^{k_1k_2k_3}\mathcal{D}_b^{k_1k_2k_3}}{E_{k_1k_2k_3}+\omega_{k_4}} \nonumber \\ 
&\hspace{1.2cm} +\frac{\mathcal{V}_b^{k_4}\mathcal{U}_a^{k_4}{\mathcal{D}}_a^{k_1k_2k_3}{\mathcal{C}}_b^{k_1k_2k_3}}{E_{k_1k_2k_3}+\omega_{k_4}}\bigg\} \nonumber \\
&=\frac{1}{4}\sum_{k_1k_2k_3k_4}\frac{\Big[\sum_a\big(\mathcal{V}_a^{k_4}\mathcal{C}_a^{k_1k_2k_3}+\mathcal{U}_a^{k_4}\mathcal{D}_a^{k_1k_2k_3}\big)\Big]^2}{E_{k_1k_2k_3k_4}}  \nonumber \\
&=\sum_{k_1k_2k_3} \ee{2}(k_1k_2k_3) \,, \label{eq:BMBPT_en} 
\end{align}
where the last equality simply amounts to isolating three-pole contributions\footnote{While not obvious at first sight, the expression in Eq.~\eqref{eq:BMBPT_en} is fully symmetric under the exchange of any pair among the four quasi-particle indices such that one can arbitrarily isolate any triplet to define the IT measure.} $(k_1k_2k_3)$ and proceeds to a trivial rewriting. The above formula thus makes explicit the relation between the correlation energy and the energy-driven IT measure introduced in Eq.~\eqref{eq:BMBPT_measure}.

\bibliographystyle{apsrev4-1}
\bibliography{biblio}

\begin{thebibliography}{57}%
\makeatletter
\providecommand \@ifxundefined [1]{%
 \@ifx{#1\undefined}
}%
\providecommand \@ifnum [1]{%
 \ifnum #1\expandafter \@firstoftwo
 \else \expandafter \@secondoftwo
 \fi
}%
\providecommand \@ifx [1]{%
 \ifx #1\expandafter \@firstoftwo
 \else \expandafter \@secondoftwo
 \fi
}%
\providecommand \natexlab [1]{#1}%
\providecommand \enquote  [1]{``#1''}%
\providecommand \bibnamefont  [1]{#1}%
\providecommand \bibfnamefont [1]{#1}%
\providecommand \citenamefont [1]{#1}%
\providecommand \href@noop [0]{\@secondoftwo}%
\providecommand \href [0]{\begingroup \@sanitize@url \@href}%
\providecommand \@href[1]{\@@startlink{#1}\@@href}%
\providecommand \@@href[1]{\endgroup#1\@@endlink}%
\providecommand \@sanitize@url [0]{\catcode `\\12\catcode `\$12\catcode
  `\&12\catcode `\#12\catcode `\^12\catcode `\_12\catcode `\%12\relax}%
\providecommand \@@startlink[1]{}%
\providecommand \@@endlink[0]{}%
\providecommand \url  [0]{\begingroup\@sanitize@url \@url }%
\providecommand \@url [1]{\endgroup\@href {#1}{\urlprefix }}%
\providecommand \urlprefix  [0]{URL }%
\providecommand \Eprint [0]{\href }%
\providecommand \doibase [0]{http://dx.doi.org/}%
\providecommand \selectlanguage [0]{\@gobble}%
\providecommand \bibinfo  [0]{\@secondoftwo}%
\providecommand \bibfield  [0]{\@secondoftwo}%
\providecommand \translation [1]{[#1]}%
\providecommand \BibitemOpen [0]{}%
\providecommand \bibitemStop [0]{}%
\providecommand \bibitemNoStop [0]{.\EOS\space}%
\providecommand \EOS [0]{\spacefactor3000\relax}%
\providecommand \BibitemShut  [1]{\csname bibitem#1\endcsname}%
\let\auto@bib@innerbib\@empty
\bibitem [{\citenamefont {Langhammer}\ \emph {et~al.}(2012)\citenamefont
  {Langhammer}, \citenamefont {Roth},\ and\ \citenamefont
  {Stumpf}}]{Langhammer:2012jx}%
  \BibitemOpen
  \bibfield  {author} {\bibinfo {author} {\bibfnamefont {J.}~\bibnamefont
  {Langhammer}}, \bibinfo {author} {\bibfnamefont {R.}~\bibnamefont {Roth}}, \
  and\ \bibinfo {author} {\bibfnamefont {C.}~\bibnamefont {Stumpf}},\ }\href
  {\doibase 10.1103/PhysRevC.86.054315} {\bibfield  {journal} {\bibinfo
  {journal} {Phys. Rev. C}\ }\textbf {\bibinfo {volume} {86}},\ \bibinfo
  {pages} {054315} (\bibinfo {year} {2012})}\BibitemShut {NoStop}%
\bibitem [{\citenamefont {Tichai}\ \emph {et~al.}(2016)\citenamefont {Tichai},
  \citenamefont {Langhammer}, \citenamefont {Binder},\ and\ \citenamefont
  {Roth}}]{Tichai2016}%
  \BibitemOpen
  \bibfield  {author} {\bibinfo {author} {\bibfnamefont {A.}~\bibnamefont
  {Tichai}}, \bibinfo {author} {\bibfnamefont {J.}~\bibnamefont {Langhammer}},
  \bibinfo {author} {\bibfnamefont {S.}~\bibnamefont {Binder}}, \ and\ \bibinfo
  {author} {\bibfnamefont {R.}~\bibnamefont {Roth}},\ }\href {\doibase
  https://doi.org/10.1016/j.physletb.2016.03.029} {\bibfield  {journal}
  {\bibinfo  {journal} {Phys. Lett. B}\ }\textbf {\bibinfo {volume} {756}},\
  \bibinfo {pages} {283} (\bibinfo {year} {2016})}\BibitemShut {NoStop}%
\bibitem [{\citenamefont {Hu}\ \emph {et~al.}(2016)\citenamefont {Hu},
  \citenamefont {Xu}, \citenamefont {Sun}, \citenamefont {Vary},\ and\
  \citenamefont {Li}}]{Hu:2016txm}%
  \BibitemOpen
  \bibfield  {author} {\bibinfo {author} {\bibfnamefont {B.}~\bibnamefont
  {Hu}}, \bibinfo {author} {\bibfnamefont {F.}~\bibnamefont {Xu}}, \bibinfo
  {author} {\bibfnamefont {Z.}~\bibnamefont {Sun}}, \bibinfo {author}
  {\bibfnamefont {J.~P.}\ \bibnamefont {Vary}}, \ and\ \bibinfo {author}
  {\bibfnamefont {T.}~\bibnamefont {Li}},\ }\href {\doibase
  10.1103/PhysRevC.94.014303} {\bibfield  {journal} {\bibinfo  {journal} {Phys.
  Rev. C}\ }\textbf {\bibinfo {volume} {94}},\ \bibinfo {pages} {014303}
  (\bibinfo {year} {2016})}\BibitemShut {NoStop}%
\bibitem [{\citenamefont {Tichai}\ \emph
  {et~al.}(2018{\natexlab{a}})\citenamefont {Tichai}, \citenamefont
  {Gebrerufael}, \citenamefont {Vobig},\ and\ \citenamefont
  {Roth}}]{Tichai:2017rqe}%
  \BibitemOpen
  \bibfield  {author} {\bibinfo {author} {\bibfnamefont {A.}~\bibnamefont
  {Tichai}}, \bibinfo {author} {\bibfnamefont {E.}~\bibnamefont {Gebrerufael}},
  \bibinfo {author} {\bibfnamefont {K.}~\bibnamefont {Vobig}}, \ and\ \bibinfo
  {author} {\bibfnamefont {R.}~\bibnamefont {Roth}},\ }\href {\doibase
  10.1016/j.physletb.2018.10.029} {\bibfield  {journal} {\bibinfo  {journal}
  {Phys. Lett. B}\ }\textbf {\bibinfo {volume} {786}},\ \bibinfo {pages} {448}
  (\bibinfo {year} {2018}{\natexlab{a}})}\BibitemShut {NoStop}%
\bibitem [{\citenamefont {Tichai}\ \emph
  {et~al.}(2018{\natexlab{b}})\citenamefont {Tichai}, \citenamefont {Arthuis},
  \citenamefont {Duguet}, \citenamefont {Hergert}, \citenamefont {Som{\`a}},\
  and\ \citenamefont {Roth}}]{Tichai:2018mll}%
  \BibitemOpen
  \bibfield  {author} {\bibinfo {author} {\bibfnamefont {A.}~\bibnamefont
  {Tichai}}, \bibinfo {author} {\bibfnamefont {P.}~\bibnamefont {Arthuis}},
  \bibinfo {author} {\bibfnamefont {T.}~\bibnamefont {Duguet}}, \bibinfo
  {author} {\bibfnamefont {H.}~\bibnamefont {Hergert}}, \bibinfo {author}
  {\bibfnamefont {V.}~\bibnamefont {Som{\`a}}}, \ and\ \bibinfo {author}
  {\bibfnamefont {R.}~\bibnamefont {Roth}},\ }\href {\doibase
  10.1016/j.physletb.2018.09.044} {\bibfield  {journal} {\bibinfo  {journal}
  {Phys. Lett. B}\ }\textbf {\bibinfo {volume} {786}},\ \bibinfo {pages} {195}
  (\bibinfo {year} {2018}{\natexlab{b}})}\BibitemShut {NoStop}%
\bibitem [{\citenamefont {Arthuis}\ \emph {et~al.}(2019)\citenamefont
  {Arthuis}, \citenamefont {Duguet}, \citenamefont {Tichai}, \citenamefont
  {Lasseri},\ and\ \citenamefont {Ebran}}]{Arthuis:2018yoo}%
  \BibitemOpen
  \bibfield  {author} {\bibinfo {author} {\bibfnamefont {P.}~\bibnamefont
  {Arthuis}}, \bibinfo {author} {\bibfnamefont {T.}~\bibnamefont {Duguet}},
  \bibinfo {author} {\bibfnamefont {A.}~\bibnamefont {Tichai}}, \bibinfo
  {author} {\bibfnamefont {R.-D.}\ \bibnamefont {Lasseri}}, \ and\ \bibinfo
  {author} {\bibfnamefont {J.-P.}\ \bibnamefont {Ebran}},\ }\href {\doibase
  10.1016/j.cpc.2018.11.023} {\bibfield  {journal} {\bibinfo  {journal}
  {Comput. Phys. Commun.}\ }\textbf {\bibinfo {volume} {240}},\ \bibinfo
  {pages} {202} (\bibinfo {year} {2019})}\BibitemShut {NoStop}%
\bibitem [{\citenamefont {Tichai}\ \emph {et~al.}(2020)\citenamefont {Tichai},
  \citenamefont {Roth},\ and\ \citenamefont {Duguet}}]{Tichai:2020dna}%
  \BibitemOpen
  \bibfield  {author} {\bibinfo {author} {\bibfnamefont {A.}~\bibnamefont
  {Tichai}}, \bibinfo {author} {\bibfnamefont {R.}~\bibnamefont {Roth}}, \ and\
  \bibinfo {author} {\bibfnamefont {T.}~\bibnamefont {Duguet}},\ }\href
  {\doibase 10.3389/fphy.2020.00164} {\bibfield  {journal} {\bibinfo  {journal}
  {Front. in Phys.}\ }\textbf {\bibinfo {volume} {8}},\ \bibinfo {pages} {164}
  (\bibinfo {year} {2020})}\BibitemShut {NoStop}%
\bibitem [{\citenamefont {Dickhoff}\ and\ \citenamefont
  {Barbieri}(2004)}]{Dickhoff:2004xx}%
  \BibitemOpen
  \bibfield  {author} {\bibinfo {author} {\bibfnamefont {W.~H.}\ \bibnamefont
  {Dickhoff}}\ and\ \bibinfo {author} {\bibfnamefont {C.}~\bibnamefont
  {Barbieri}},\ }\href {\doibase 10.1016/j.ppnp.2004.02.038} {\bibfield
  {journal} {\bibinfo  {journal} {Prog. Part. Nucl. Phys.}\ }\textbf {\bibinfo
  {volume} {52}},\ \bibinfo {pages} {377} (\bibinfo {year} {2004})}\BibitemShut
  {NoStop}%
\bibitem [{\citenamefont {Som{\`a}}\ \emph {et~al.}(2011)\citenamefont
  {Som{\`a}}, \citenamefont {Duguet},\ and\ \citenamefont
  {Barbieri}}]{Soma:2011aj}%
  \BibitemOpen
  \bibfield  {author} {\bibinfo {author} {\bibfnamefont {V.}~\bibnamefont
  {Som{\`a}}}, \bibinfo {author} {\bibfnamefont {T.}~\bibnamefont {Duguet}}, \
  and\ \bibinfo {author} {\bibfnamefont {C.}~\bibnamefont {Barbieri}},\ }\href
  {\doibase 10.1103/PhysRevC.84.064317} {\bibfield  {journal} {\bibinfo
  {journal} {Phys. Rev. C}\ }\textbf {\bibinfo {volume} {84}},\ \bibinfo
  {pages} {064317} (\bibinfo {year} {2011})}\BibitemShut {NoStop}%
\bibitem [{\citenamefont {Carbone}\ \emph {et~al.}(2013)\citenamefont
  {Carbone}, \citenamefont {Cipollone}, \citenamefont {Barbieri}, \citenamefont
  {Rios},\ and\ \citenamefont {Polls}}]{Carbone:2013eqa}%
  \BibitemOpen
  \bibfield  {author} {\bibinfo {author} {\bibfnamefont {A.}~\bibnamefont
  {Carbone}}, \bibinfo {author} {\bibfnamefont {A.}~\bibnamefont {Cipollone}},
  \bibinfo {author} {\bibfnamefont {C.}~\bibnamefont {Barbieri}}, \bibinfo
  {author} {\bibfnamefont {A.}~\bibnamefont {Rios}}, \ and\ \bibinfo {author}
  {\bibfnamefont {A.}~\bibnamefont {Polls}},\ }\href {\doibase
  10.1103/PhysRevC.88.054326} {\bibfield  {journal} {\bibinfo  {journal} {Phys.
  Rev. C}\ }\textbf {\bibinfo {volume} {88}},\ \bibinfo {pages} {054326}
  (\bibinfo {year} {2013})}\BibitemShut {NoStop}%
\bibitem [{\citenamefont {Som{\`a}}\ \emph {et~al.}(2014)\citenamefont
  {Som{\`a}}, \citenamefont {Cipollone}, \citenamefont {Barbieri},
  \citenamefont {Navr{\'a}til},\ and\ \citenamefont {Duguet}}]{Soma:2013xha}%
  \BibitemOpen
  \bibfield  {author} {\bibinfo {author} {\bibfnamefont {V.}~\bibnamefont
  {Som{\`a}}}, \bibinfo {author} {\bibfnamefont {A.}~\bibnamefont {Cipollone}},
  \bibinfo {author} {\bibfnamefont {C.}~\bibnamefont {Barbieri}}, \bibinfo
  {author} {\bibfnamefont {P.}~\bibnamefont {Navr{\'a}til}}, \ and\ \bibinfo
  {author} {\bibfnamefont {T.}~\bibnamefont {Duguet}},\ }\href {\doibase
  10.1103/PhysRevC.89.061301} {\bibfield  {journal} {\bibinfo  {journal} {Phys.
  Rev. C}\ }\textbf {\bibinfo {volume} {89}},\ \bibinfo {pages} {061301}
  (\bibinfo {year} {2014})}\BibitemShut {NoStop}%
\bibitem [{\citenamefont {Lapoux}\ \emph {et~al.}(2016)\citenamefont {Lapoux},
  \citenamefont {Som{\`a}}, \citenamefont {Barbieri}, \citenamefont {Hergert},
  \citenamefont {Holt},\ and\ \citenamefont {Stroberg}}]{Lapoux:2016exf}%
  \BibitemOpen
  \bibfield  {author} {\bibinfo {author} {\bibfnamefont {V.}~\bibnamefont
  {Lapoux}}, \bibinfo {author} {\bibfnamefont {V.}~\bibnamefont {Som{\`a}}},
  \bibinfo {author} {\bibfnamefont {C.}~\bibnamefont {Barbieri}}, \bibinfo
  {author} {\bibfnamefont {H.}~\bibnamefont {Hergert}}, \bibinfo {author}
  {\bibfnamefont {J.~D.}\ \bibnamefont {Holt}}, \ and\ \bibinfo {author}
  {\bibfnamefont {S.}~\bibnamefont {Stroberg}},\ }\href {\doibase
  10.1103/PhysRevLett.117.052501} {\bibfield  {journal} {\bibinfo  {journal}
  {Phys. Rev. Lett.}\ }\textbf {\bibinfo {volume} {117}},\ \bibinfo {pages}
  {052501} (\bibinfo {year} {2016})}\BibitemShut {NoStop}%
\bibitem [{\citenamefont {Duguet}\ \emph {et~al.}(2017)\citenamefont {Duguet},
  \citenamefont {Som{\`a}}, \citenamefont {Lecluse}, \citenamefont {Barbieri},\
  and\ \citenamefont {Navr{\'a}til}}]{Duguet:2016wwr}%
  \BibitemOpen
  \bibfield  {author} {\bibinfo {author} {\bibfnamefont {T.}~\bibnamefont
  {Duguet}}, \bibinfo {author} {\bibfnamefont {V.}~\bibnamefont {Som{\`a}}},
  \bibinfo {author} {\bibfnamefont {S.}~\bibnamefont {Lecluse}}, \bibinfo
  {author} {\bibfnamefont {C.}~\bibnamefont {Barbieri}}, \ and\ \bibinfo
  {author} {\bibfnamefont {P.}~\bibnamefont {Navr{\'a}til}},\ }\href {\doibase
  10.1103/PhysRevC.95.034319} {\bibfield  {journal} {\bibinfo  {journal} {Phys.
  Rev. C}\ }\textbf {\bibinfo {volume} {95}},\ \bibinfo {pages} {034319}
  (\bibinfo {year} {2017})}\BibitemShut {NoStop}%
\bibitem [{\citenamefont {Raimondi}\ and\ \citenamefont
  {Barbieri}(2019)}]{Raimondi:2018mtv}%
  \BibitemOpen
  \bibfield  {author} {\bibinfo {author} {\bibfnamefont {F.}~\bibnamefont
  {Raimondi}}\ and\ \bibinfo {author} {\bibfnamefont {C.}~\bibnamefont
  {Barbieri}},\ }\href {\doibase 10.1103/PhysRevC.99.054327} {\bibfield
  {journal} {\bibinfo  {journal} {Phys. Rev. C}\ }\textbf {\bibinfo {volume}
  {99}},\ \bibinfo {pages} {054327} (\bibinfo {year} {2019})}\BibitemShut
  {NoStop}%
\bibitem [{\citenamefont {Arthuis}\ \emph {et~al.}(2020)\citenamefont
  {Arthuis}, \citenamefont {Barbieri}, \citenamefont {Vorabbi},\ and\
  \citenamefont {Finelli}}]{Arthuis20}%
  \BibitemOpen
  \bibfield  {author} {\bibinfo {author} {\bibfnamefont {P.}~\bibnamefont
  {Arthuis}}, \bibinfo {author} {\bibfnamefont {C.}~\bibnamefont {Barbieri}},
  \bibinfo {author} {\bibfnamefont {M.}~\bibnamefont {Vorabbi}}, \ and\
  \bibinfo {author} {\bibfnamefont {P.}~\bibnamefont {Finelli}},\ }\href
  {\doibase 10.1103/PhysRevLett.125.182501} {\bibfield  {journal} {\bibinfo
  {journal} {Phys. Rev. Lett.}\ }\textbf {\bibinfo {volume} {125}},\ \bibinfo
  {pages} {182501} (\bibinfo {year} {2020})}\BibitemShut {NoStop}%
\bibitem [{\citenamefont {Som\`a}(2020)}]{Soma20b}%
  \BibitemOpen
  \bibfield  {author} {\bibinfo {author} {\bibfnamefont {V.}~\bibnamefont
  {Som\`a}},\ }\href {\doibase 10.3389/fphy.2020.00340} {\bibfield  {journal}
  {\bibinfo  {journal} {Front. in Phys.}\ }\textbf {\bibinfo {volume} {8}},\
  \bibinfo {pages} {340} (\bibinfo {year} {2020})}\BibitemShut {NoStop}%
\bibitem [{\citenamefont {Hagen}\ \emph {et~al.}(2012)\citenamefont {Hagen},
  \citenamefont {Hjorth-Jensen}, \citenamefont {Jansen}, \citenamefont
  {Machleidt},\ and\ \citenamefont {Papenbrock}}]{Hagen:2012sh}%
  \BibitemOpen
  \bibfield  {author} {\bibinfo {author} {\bibfnamefont {G.}~\bibnamefont
  {Hagen}}, \bibinfo {author} {\bibfnamefont {M.}~\bibnamefont
  {Hjorth-Jensen}}, \bibinfo {author} {\bibfnamefont {G.}~\bibnamefont
  {Jansen}}, \bibinfo {author} {\bibfnamefont {R.}~\bibnamefont {Machleidt}}, \
  and\ \bibinfo {author} {\bibfnamefont {T.}~\bibnamefont {Papenbrock}},\
  }\href {\doibase 10.1103/PhysRevLett.108.242501} {\bibfield  {journal}
  {\bibinfo  {journal} {Phys. Rev. Lett.}\ }\textbf {\bibinfo {volume} {108}},\
  \bibinfo {pages} {242501} (\bibinfo {year} {2012})}\BibitemShut {NoStop}%
\bibitem [{\citenamefont {Binder}\ \emph {et~al.}(2013)\citenamefont {Binder},
  \citenamefont {Piecuch}, \citenamefont {Calci}, \citenamefont {Langhammer},
  \citenamefont {Navr\'atil},\ and\ \citenamefont {Roth}}]{Binder:2013oea}%
  \BibitemOpen
  \bibfield  {author} {\bibinfo {author} {\bibfnamefont {S.}~\bibnamefont
  {Binder}}, \bibinfo {author} {\bibfnamefont {P.}~\bibnamefont {Piecuch}},
  \bibinfo {author} {\bibfnamefont {A.}~\bibnamefont {Calci}}, \bibinfo
  {author} {\bibfnamefont {J.}~\bibnamefont {Langhammer}}, \bibinfo {author}
  {\bibfnamefont {P.}~\bibnamefont {Navr\'atil}}, \ and\ \bibinfo {author}
  {\bibfnamefont {R.}~\bibnamefont {Roth}},\ }\href {\doibase
  10.1103/PhysRevC.88.054319} {\bibfield  {journal} {\bibinfo  {journal} {Phys.
  Rev. C}\ }\textbf {\bibinfo {volume} {88}},\ \bibinfo {pages} {054319}
  (\bibinfo {year} {2013})}\BibitemShut {NoStop}%
\bibitem [{\citenamefont {Binder}\ \emph {et~al.}(2014)\citenamefont {Binder},
  \citenamefont {Langhammer}, \citenamefont {Calci},\ and\ \citenamefont
  {Roth}}]{Binder:2013xaa}%
  \BibitemOpen
  \bibfield  {author} {\bibinfo {author} {\bibfnamefont {S.}~\bibnamefont
  {Binder}}, \bibinfo {author} {\bibfnamefont {J.}~\bibnamefont {Langhammer}},
  \bibinfo {author} {\bibfnamefont {A.}~\bibnamefont {Calci}}, \ and\ \bibinfo
  {author} {\bibfnamefont {R.}~\bibnamefont {Roth}},\ }\href {\doibase
  10.1016/j.physletb.2014.07.010} {\bibfield  {journal} {\bibinfo  {journal}
  {Phys. Lett. B}\ }\textbf {\bibinfo {volume} {736}},\ \bibinfo {pages} {119}
  (\bibinfo {year} {2014})}\BibitemShut {NoStop}%
\bibitem [{\citenamefont {Hagen}\ \emph {et~al.}(2014)\citenamefont {Hagen},
  \citenamefont {Papenbrock}, \citenamefont {Hjorth-Jensen},\ and\
  \citenamefont {Dean}}]{Hagen:2013nca}%
  \BibitemOpen
  \bibfield  {author} {\bibinfo {author} {\bibfnamefont {G.}~\bibnamefont
  {Hagen}}, \bibinfo {author} {\bibfnamefont {T.}~\bibnamefont {Papenbrock}},
  \bibinfo {author} {\bibfnamefont {M.}~\bibnamefont {Hjorth-Jensen}}, \ and\
  \bibinfo {author} {\bibfnamefont {D.~J.}\ \bibnamefont {Dean}},\ }\href
  {\doibase 10.1088/0034-4885/77/9/096302} {\bibfield  {journal} {\bibinfo
  {journal} {Rept. Prog. Phys.}\ }\textbf {\bibinfo {volume} {77}},\ \bibinfo
  {pages} {096302} (\bibinfo {year} {2014})}\BibitemShut {NoStop}%
\bibitem [{\citenamefont {Signoracci}\ \emph {et~al.}(2015)\citenamefont
  {Signoracci}, \citenamefont {Duguet}, \citenamefont {Hagen},\ and\
  \citenamefont {Jansen}}]{Signoracci:2014dia}%
  \BibitemOpen
  \bibfield  {author} {\bibinfo {author} {\bibfnamefont {A.}~\bibnamefont
  {Signoracci}}, \bibinfo {author} {\bibfnamefont {T.}~\bibnamefont {Duguet}},
  \bibinfo {author} {\bibfnamefont {G.}~\bibnamefont {Hagen}}, \ and\ \bibinfo
  {author} {\bibfnamefont {G.}~\bibnamefont {Jansen}},\ }\href {\doibase
  10.1103/PhysRevC.91.064320} {\bibfield  {journal} {\bibinfo  {journal} {Phys.
  Rev. C}\ }\textbf {\bibinfo {volume} {91}},\ \bibinfo {pages} {064320}
  (\bibinfo {year} {2015})}\BibitemShut {NoStop}%
\bibitem [{\citenamefont {Morris}\ \emph {et~al.}(2018)\citenamefont {Morris},
  \citenamefont {Simonis}, \citenamefont {Stroberg}, \citenamefont {Stumpf},
  \citenamefont {Hagen}, \citenamefont {Holt}, \citenamefont {Jansen},
  \citenamefont {Papenbrock}, \citenamefont {Roth},\ and\ \citenamefont
  {Schwenk}}]{Morris:2017vxi}%
  \BibitemOpen
  \bibfield  {author} {\bibinfo {author} {\bibfnamefont {T.~D.}\ \bibnamefont
  {Morris}}, \bibinfo {author} {\bibfnamefont {J.}~\bibnamefont {Simonis}},
  \bibinfo {author} {\bibfnamefont {S.~R.}\ \bibnamefont {Stroberg}}, \bibinfo
  {author} {\bibfnamefont {C.}~\bibnamefont {Stumpf}}, \bibinfo {author}
  {\bibfnamefont {G.}~\bibnamefont {Hagen}}, \bibinfo {author} {\bibfnamefont
  {J.~D.}\ \bibnamefont {Holt}}, \bibinfo {author} {\bibfnamefont {G.~R.}\
  \bibnamefont {Jansen}}, \bibinfo {author} {\bibfnamefont {T.}~\bibnamefont
  {Papenbrock}}, \bibinfo {author} {\bibfnamefont {R.}~\bibnamefont {Roth}}, \
  and\ \bibinfo {author} {\bibfnamefont {A.}~\bibnamefont {Schwenk}},\ }\href
  {\doibase 10.1103/PhysRevLett.120.152503} {\bibfield  {journal} {\bibinfo
  {journal} {Phys. Rev. Lett.}\ }\textbf {\bibinfo {volume} {120}},\ \bibinfo
  {pages} {152503} (\bibinfo {year} {2018})}\BibitemShut {NoStop}%
\bibitem [{\citenamefont {Tsukiyama}\ \emph {et~al.}(2011)\citenamefont
  {Tsukiyama}, \citenamefont {Bogner},\ and\ \citenamefont
  {Schwenk}}]{Tsukiyama2011prl}%
  \BibitemOpen
  \bibfield  {author} {\bibinfo {author} {\bibfnamefont {K.}~\bibnamefont
  {Tsukiyama}}, \bibinfo {author} {\bibfnamefont {S.~K.}\ \bibnamefont
  {Bogner}}, \ and\ \bibinfo {author} {\bibfnamefont {A.}~\bibnamefont
  {Schwenk}},\ }\href {\doibase 10.1103/PhysRevLett.106.222502} {\bibfield
  {journal} {\bibinfo  {journal} {Phys. Rev. Lett.}\ }\textbf {\bibinfo
  {volume} {106}},\ \bibinfo {pages} {222502} (\bibinfo {year}
  {2011})}\BibitemShut {NoStop}%
\bibitem [{\citenamefont {Hergert}\ \emph {et~al.}(2016)\citenamefont
  {Hergert}, \citenamefont {Bogner}, \citenamefont {Morris}, \citenamefont
  {Schwenk},\ and\ \citenamefont {Tsukiyama}}]{Hergert:2015awm}%
  \BibitemOpen
  \bibfield  {author} {\bibinfo {author} {\bibfnamefont {H.}~\bibnamefont
  {Hergert}}, \bibinfo {author} {\bibfnamefont {S.~K.}\ \bibnamefont {Bogner}},
  \bibinfo {author} {\bibfnamefont {T.~D.}\ \bibnamefont {Morris}}, \bibinfo
  {author} {\bibfnamefont {A.}~\bibnamefont {Schwenk}}, \ and\ \bibinfo
  {author} {\bibfnamefont {K.}~\bibnamefont {Tsukiyama}},\ }\href {\doibase
  10.1016/j.physrep.2015.12.007} {\bibfield  {journal} {\bibinfo  {journal}
  {Phys. Rept.}\ }\textbf {\bibinfo {volume} {621}},\ \bibinfo {pages} {165}
  (\bibinfo {year} {2016})}\BibitemShut {NoStop}%
\bibitem [{\citenamefont {Hergert}\ \emph {et~al.}(2017)\citenamefont
  {Hergert}, \citenamefont {Bogner}, \citenamefont {Lietz}, \citenamefont
  {Morris}, \citenamefont {Novario}, \citenamefont {Parzuchowski},\ and\
  \citenamefont {Yuan}}]{Hergert:2016iju}%
  \BibitemOpen
  \bibfield  {author} {\bibinfo {author} {\bibfnamefont {H.}~\bibnamefont
  {Hergert}}, \bibinfo {author} {\bibfnamefont {S.~K.}\ \bibnamefont {Bogner}},
  \bibinfo {author} {\bibfnamefont {J.~G.}\ \bibnamefont {Lietz}}, \bibinfo
  {author} {\bibfnamefont {T.~D.}\ \bibnamefont {Morris}}, \bibinfo {author}
  {\bibfnamefont {S.}~\bibnamefont {Novario}}, \bibinfo {author} {\bibfnamefont
  {N.~M.}\ \bibnamefont {Parzuchowski}}, \ and\ \bibinfo {author}
  {\bibfnamefont {F.}~\bibnamefont {Yuan}},\ }\href {\doibase
  10.1007/978-3-319-53336-0_10} {\bibfield  {journal} {\bibinfo  {journal}
  {Lect. Notes Phys.}\ }\textbf {\bibinfo {volume} {936}},\ \bibinfo {pages}
  {477} (\bibinfo {year} {2017})}\BibitemShut {NoStop}%
\bibitem [{\citenamefont {Parzuchowski}\ \emph {et~al.}(2017)\citenamefont
  {Parzuchowski}, \citenamefont {Stroberg}, \citenamefont {Navr{\'a}til},
  \citenamefont {Hergert},\ and\ \citenamefont
  {Bogner}}]{Parzuchowski:2017wcq}%
  \BibitemOpen
  \bibfield  {author} {\bibinfo {author} {\bibfnamefont {N.~M.}\ \bibnamefont
  {Parzuchowski}}, \bibinfo {author} {\bibfnamefont {S.~R.}\ \bibnamefont
  {Stroberg}}, \bibinfo {author} {\bibfnamefont {P.}~\bibnamefont
  {Navr{\'a}til}}, \bibinfo {author} {\bibfnamefont {H.}~\bibnamefont
  {Hergert}}, \ and\ \bibinfo {author} {\bibfnamefont {S.~K.}\ \bibnamefont
  {Bogner}},\ }\href {\doibase 10.1103/PhysRevC.96.034324} {\bibfield
  {journal} {\bibinfo  {journal} {Phys. Rev. C}\ }\textbf {\bibinfo {volume}
  {96}},\ \bibinfo {pages} {034324} (\bibinfo {year} {2017})}\BibitemShut
  {NoStop}%
\bibitem [{\citenamefont {Heinz}\ \emph {et~al.}(2021)\citenamefont {Heinz},
  \citenamefont {Tichai}, \citenamefont {Hoppe}, \citenamefont {Hebeler},\ and\
  \citenamefont {Schwenk}}]{Heinz2021}%
  \BibitemOpen
  \bibfield  {author} {\bibinfo {author} {\bibfnamefont {M.}~\bibnamefont
  {Heinz}}, \bibinfo {author} {\bibfnamefont {A.}~\bibnamefont {Tichai}},
  \bibinfo {author} {\bibfnamefont {J.}~\bibnamefont {Hoppe}}, \bibinfo
  {author} {\bibfnamefont {K.}~\bibnamefont {Hebeler}}, \ and\ \bibinfo
  {author} {\bibfnamefont {A.}~\bibnamefont {Schwenk}},\ }\href@noop {} {\
  (\bibinfo {year} {2021})},\ \Eprint {http://arxiv.org/abs/2102.11172}
  {arXiv:2102.11172 [nucl-th]} \BibitemShut {NoStop}%
\bibitem [{\citenamefont {Duguet}(2015)}]{Duguet:2014jja}%
  \BibitemOpen
  \bibfield  {author} {\bibinfo {author} {\bibfnamefont {T.}~\bibnamefont
  {Duguet}},\ }\href {\doibase 10.1088/0954-3899/42/2/025107} {\bibfield
  {journal} {\bibinfo  {journal} {J. Phys. G: Nucl. Part. Phys.}\ }\textbf
  {\bibinfo {volume} {42}},\ \bibinfo {pages} {025107} (\bibinfo {year}
  {2015})}\BibitemShut {NoStop}%
\bibitem [{\citenamefont {Duguet}\ and\ \citenamefont
  {Signoracci}(2017)}]{Duguet:2015yle}%
  \BibitemOpen
  \bibfield  {author} {\bibinfo {author} {\bibfnamefont {T.}~\bibnamefont
  {Duguet}}\ and\ \bibinfo {author} {\bibfnamefont {A.}~\bibnamefont
  {Signoracci}},\ }\href {\doibase 10.1088/1361-6471/aa5d3e} {\bibfield
  {journal} {\bibinfo  {journal} {J. Phys. G: Nucl. Part. Phys.}\ }\textbf
  {\bibinfo {volume} {44}},\ \bibinfo {pages} {015103} (\bibinfo {year}
  {2017})}\BibitemShut {NoStop}%
\bibitem [{\citenamefont {Qiu}\ \emph {et~al.}(2017)\citenamefont {Qiu},
  \citenamefont {Henderson}, \citenamefont {Zhao},\ and\ \citenamefont
  {Scuseria}}]{qiu17a}%
  \BibitemOpen
  \bibfield  {author} {\bibinfo {author} {\bibfnamefont {Y.}~\bibnamefont
  {Qiu}}, \bibinfo {author} {\bibfnamefont {T.~M.}\ \bibnamefont {Henderson}},
  \bibinfo {author} {\bibfnamefont {J.}~\bibnamefont {Zhao}}, \ and\ \bibinfo
  {author} {\bibfnamefont {G.~E.}\ \bibnamefont {Scuseria}},\ }\href {\doibase
  10.1063/1.4991020} {\bibfield  {journal} {\bibinfo  {journal} {J. Chem.
  Phys.}\ }\textbf {\bibinfo {volume} {147}},\ \bibinfo {pages} {064111}
  (\bibinfo {year} {2017})}\BibitemShut {NoStop}%
\bibitem [{\citenamefont {Qiu}\ \emph {et~al.}(2019)\citenamefont {Qiu},
  \citenamefont {Henderson}, \citenamefont {Duguet},\ and\ \citenamefont
  {Scuseria}}]{Qiu:2018edx}%
  \BibitemOpen
  \bibfield  {author} {\bibinfo {author} {\bibfnamefont {Y.}~\bibnamefont
  {Qiu}}, \bibinfo {author} {\bibfnamefont {T.}~\bibnamefont {Henderson}},
  \bibinfo {author} {\bibfnamefont {T.}~\bibnamefont {Duguet}}, \ and\ \bibinfo
  {author} {\bibfnamefont {G.}~\bibnamefont {Scuseria}},\ }\href {\doibase
  10.1103/PhysRevC.99.044301} {\bibfield  {journal} {\bibinfo  {journal} {Phys.
  Rev. C}\ }\textbf {\bibinfo {volume} {99}},\ \bibinfo {pages} {044301}
  (\bibinfo {year} {2019})}\BibitemShut {NoStop}%
\bibitem [{\citenamefont {Roth}(2009)}]{Roth2009}%
  \BibitemOpen
  \bibfield  {author} {\bibinfo {author} {\bibfnamefont {R.}~\bibnamefont
  {Roth}},\ }\href {\doibase 10.1103/PhysRevC.79.064324} {\bibfield  {journal}
  {\bibinfo  {journal} {Phys. Rev. C}\ }\textbf {\bibinfo {volume} {79}},\
  \bibinfo {pages} {064324} (\bibinfo {year} {2009})}\BibitemShut {NoStop}%
\bibitem [{\citenamefont {Roth}\ and\ \citenamefont
  {Navr{\'a}til}(2007)}]{roth2007ab}%
  \BibitemOpen
  \bibfield  {author} {\bibinfo {author} {\bibfnamefont {R.}~\bibnamefont
  {Roth}}\ and\ \bibinfo {author} {\bibfnamefont {P.}~\bibnamefont
  {Navr{\'a}til}},\ }\href {\doibase 10.1103/PhysRevLett.99.092501} {\bibfield
  {journal} {\bibinfo  {journal} {Phys. Rev. Lett.}\ }\textbf {\bibinfo
  {volume} {99}},\ \bibinfo {pages} {092501} (\bibinfo {year}
  {2007})}\BibitemShut {NoStop}%
\bibitem [{\citenamefont {Tichai}\ \emph {et~al.}(2019)\citenamefont {Tichai},
  \citenamefont {Ripoche},\ and\ \citenamefont {Duguet}}]{tichai2019pre}%
  \BibitemOpen
  \bibfield  {author} {\bibinfo {author} {\bibfnamefont {A.}~\bibnamefont
  {Tichai}}, \bibinfo {author} {\bibfnamefont {J.}~\bibnamefont {Ripoche}}, \
  and\ \bibinfo {author} {\bibfnamefont {T.}~\bibnamefont {Duguet}},\ }\href
  {\doibase 10.1140/epja/i2019-12758-6} {\bibfield  {journal} {\bibinfo
  {journal} {Eur. Phys. J. A}\ }\textbf {\bibinfo {volume} {55}},\ \bibinfo
  {pages} {90} (\bibinfo {year} {2019})}\BibitemShut {NoStop}%
\bibitem [{\citenamefont {Som\`a}\ \emph {et~al.}(2014)\citenamefont {Som\`a},
  \citenamefont {Barbieri},\ and\ \citenamefont {Duguet}}]{Soma:2013ona}%
  \BibitemOpen
  \bibfield  {author} {\bibinfo {author} {\bibfnamefont {V.}~\bibnamefont
  {Som\`a}}, \bibinfo {author} {\bibfnamefont {C.}~\bibnamefont {Barbieri}}, \
  and\ \bibinfo {author} {\bibfnamefont {T.}~\bibnamefont {Duguet}},\ }\href
  {\doibase 10.1103/PhysRevC.89.024323} {\bibfield  {journal} {\bibinfo
  {journal} {Phys. Rev. C}\ }\textbf {\bibinfo {volume} {89}},\ \bibinfo
  {pages} {024323} (\bibinfo {year} {2014})}\BibitemShut {NoStop}%
\bibitem [{\citenamefont {Ripoche}\ \emph {et~al.}(2020)\citenamefont
  {Ripoche}, \citenamefont {Tichai},\ and\ \citenamefont
  {Duguet}}]{Ripoche:2019nmy}%
  \BibitemOpen
  \bibfield  {author} {\bibinfo {author} {\bibfnamefont {J.}~\bibnamefont
  {Ripoche}}, \bibinfo {author} {\bibfnamefont {A.}~\bibnamefont {Tichai}}, \
  and\ \bibinfo {author} {\bibfnamefont {T.}~\bibnamefont {Duguet}},\ }\href
  {\doibase 10.1140/epja/s10050-020-00045-8} {\bibfield  {journal} {\bibinfo
  {journal} {Eur. Phys. J. A}\ }\textbf {\bibinfo {volume} {56}},\ \bibinfo
  {pages} {40} (\bibinfo {year} {2020})}\BibitemShut {NoStop}%
\bibitem [{\citenamefont {Frosini}\ \emph {et~al.}(2021)\citenamefont
  {Frosini}, \citenamefont {Duguet}, \citenamefont {Bally}, \citenamefont
  {Beaujeault-Taudi\`ere}, \citenamefont {Ebran},\ and\ \citenamefont
  {Som\`a}}]{Frosini:2021tuj}%
  \BibitemOpen
  \bibfield  {author} {\bibinfo {author} {\bibfnamefont {M.}~\bibnamefont
  {Frosini}}, \bibinfo {author} {\bibfnamefont {T.}~\bibnamefont {Duguet}},
  \bibinfo {author} {\bibfnamefont {B.}~\bibnamefont {Bally}}, \bibinfo
  {author} {\bibfnamefont {Y.}~\bibnamefont {Beaujeault-Taudi\`ere}}, \bibinfo
  {author} {\bibfnamefont {J.~P.}\ \bibnamefont {Ebran}}, \ and\ \bibinfo
  {author} {\bibfnamefont {V.}~\bibnamefont {Som\`a}},\ }\href@noop {} {\
  (\bibinfo {year} {2021})},\ \Eprint {http://arxiv.org/abs/2102.10120}
  {arXiv:2102.10120 [nucl-th]} \BibitemShut {NoStop}%
\bibitem [{\citenamefont {Ring}\ and\ \citenamefont {Schuck}(2005)}]{Ring}%
  \BibitemOpen
  \bibfield  {author} {\bibinfo {author} {\bibfnamefont {P.}~\bibnamefont
  {Ring}}\ and\ \bibinfo {author} {\bibfnamefont {P.}~\bibnamefont {Schuck}},\
  }\href {https://www.xarg.org/ref/a/354021206X/} {\emph {\bibinfo {title} {The
  Nuclear Many-Body Problem (Theoretical and Mathematical Physics)}}}\
  (\bibinfo  {publisher} {Springer},\ \bibinfo {year} {2005})\BibitemShut
  {NoStop}%
\bibitem [{\citenamefont {Schirmer}\ \emph {et~al.}(1983)\citenamefont
  {Schirmer}, \citenamefont {Cederbaum},\ and\ \citenamefont
  {Walter}}]{Schirmer83}%
  \BibitemOpen
  \bibfield  {author} {\bibinfo {author} {\bibfnamefont {J.}~\bibnamefont
  {Schirmer}}, \bibinfo {author} {\bibfnamefont {L.~S.}\ \bibnamefont
  {Cederbaum}}, \ and\ \bibinfo {author} {\bibfnamefont {O.}~\bibnamefont
  {Walter}},\ }\href {\doibase 10.1103/PhysRevA.28.1237} {\bibfield  {journal}
  {\bibinfo  {journal} {Phys. Rev. A}\ }\textbf {\bibinfo {volume} {28}},\
  \bibinfo {pages} {1237} (\bibinfo {year} {1983})}\BibitemShut {NoStop}%
\bibitem [{\citenamefont {Cipollone}\ \emph {et~al.}(2013)\citenamefont
  {Cipollone}, \citenamefont {Barbieri},\ and\ \citenamefont
  {Navr{\'a}til}}]{cipollone2013isotopic}%
  \BibitemOpen
  \bibfield  {author} {\bibinfo {author} {\bibfnamefont {A.}~\bibnamefont
  {Cipollone}}, \bibinfo {author} {\bibfnamefont {C.}~\bibnamefont {Barbieri}},
  \ and\ \bibinfo {author} {\bibfnamefont {P.}~\bibnamefont {Navr{\'a}til}},\
  }\href {\doibase 10.1103/PhysRevLett.111.062501} {\bibfield  {journal}
  {\bibinfo  {journal} {Phys. Rev. Lett.}\ }\textbf {\bibinfo {volume} {111}},\
  \bibinfo {pages} {062501} (\bibinfo {year} {2013})}\BibitemShut {NoStop}%
\bibitem [{\citenamefont {Raimondi}\ and\ \citenamefont
  {Barbieri}(2018)}]{raimondi2018}%
  \BibitemOpen
  \bibfield  {author} {\bibinfo {author} {\bibfnamefont {F.}~\bibnamefont
  {Raimondi}}\ and\ \bibinfo {author} {\bibfnamefont {C.}~\bibnamefont
  {Barbieri}},\ }\href {\doibase 10.1103/PhysRevC.97.054308} {\bibfield
  {journal} {\bibinfo  {journal} {Phys. Rev. C}\ }\textbf {\bibinfo {volume}
  {97}},\ \bibinfo {pages} {054308} (\bibinfo {year} {2018})}\BibitemShut
  {NoStop}%
\bibitem [{\citenamefont {Barbieri}\ and\ \citenamefont
  {Hjorth-Jensen}(2009)}]{Barbieri09}%
  \BibitemOpen
  \bibfield  {author} {\bibinfo {author} {\bibfnamefont {C.}~\bibnamefont
  {Barbieri}}\ and\ \bibinfo {author} {\bibfnamefont {M.}~\bibnamefont
  {Hjorth-Jensen}},\ }\href {\doibase 10.1103/PhysRevC.79.064313} {\bibfield
  {journal} {\bibinfo  {journal} {Phys. Rev. C}\ }\textbf {\bibinfo {volume}
  {79}},\ \bibinfo {pages} {064313} (\bibinfo {year} {2009})}\BibitemShut
  {NoStop}%
\bibitem [{\citenamefont {Som\`a}\ \emph
  {et~al.}(2020{\natexlab{a}})\citenamefont {Som\`a}, \citenamefont
  {Navr\'atil}, \citenamefont {Raimondi}, \citenamefont {Barbieri},\ and\
  \citenamefont {Duguet}}]{Soma20}%
  \BibitemOpen
  \bibfield  {author} {\bibinfo {author} {\bibfnamefont {V.}~\bibnamefont
  {Som\`a}}, \bibinfo {author} {\bibfnamefont {P.}~\bibnamefont {Navr\'atil}},
  \bibinfo {author} {\bibfnamefont {F.}~\bibnamefont {Raimondi}}, \bibinfo
  {author} {\bibfnamefont {C.}~\bibnamefont {Barbieri}}, \ and\ \bibinfo
  {author} {\bibfnamefont {T.}~\bibnamefont {Duguet}},\ }\href {\doibase
  10.1103/PhysRevC.101.014318} {\bibfield  {journal} {\bibinfo  {journal}
  {Phys. Rev. C}\ }\textbf {\bibinfo {volume} {101}},\ \bibinfo {pages}
  {014318} (\bibinfo {year} {2020}{\natexlab{a}})}\BibitemShut {NoStop}%
\bibitem [{\citenamefont {Cipollone}\ \emph {et~al.}(2015)\citenamefont
  {Cipollone}, \citenamefont {Barbieri},\ and\ \citenamefont
  {Navr{\'a}til}}]{cipollone2015chiral}%
  \BibitemOpen
  \bibfield  {author} {\bibinfo {author} {\bibfnamefont {A.}~\bibnamefont
  {Cipollone}}, \bibinfo {author} {\bibfnamefont {C.}~\bibnamefont {Barbieri}},
  \ and\ \bibinfo {author} {\bibfnamefont {P.}~\bibnamefont {Navr{\'a}til}},\
  }\href {\doibase 10.1103/PhysRevC.92.014306} {\bibfield  {journal} {\bibinfo
  {journal} {Phys. Rev. C}\ }\textbf {\bibinfo {volume} {92}},\ \bibinfo
  {pages} {014306} (\bibinfo {year} {2015})}\BibitemShut {NoStop}%
\bibitem [{\citenamefont {Ekstr\"om}\ \emph {et~al.}(2015)\citenamefont
  {Ekstr\"om}, \citenamefont {Jansen}, \citenamefont {Wendt}, \citenamefont
  {Hagen}, \citenamefont {Papenbrock}, \citenamefont {Carlsson}, \citenamefont
  {Forss\'en}, \citenamefont {Hjorth-Jensen}, \citenamefont {Navr\'atil},\ and\
  \citenamefont {Nazarewicz}}]{Ekstrom15}%
  \BibitemOpen
  \bibfield  {author} {\bibinfo {author} {\bibfnamefont {A.}~\bibnamefont
  {Ekstr\"om}}, \bibinfo {author} {\bibfnamefont {G.~R.}\ \bibnamefont
  {Jansen}}, \bibinfo {author} {\bibfnamefont {K.~A.}\ \bibnamefont {Wendt}},
  \bibinfo {author} {\bibfnamefont {G.}~\bibnamefont {Hagen}}, \bibinfo
  {author} {\bibfnamefont {T.}~\bibnamefont {Papenbrock}}, \bibinfo {author}
  {\bibfnamefont {B.~D.}\ \bibnamefont {Carlsson}}, \bibinfo {author}
  {\bibfnamefont {C.}~\bibnamefont {Forss\'en}}, \bibinfo {author}
  {\bibfnamefont {M.}~\bibnamefont {Hjorth-Jensen}}, \bibinfo {author}
  {\bibfnamefont {P.}~\bibnamefont {Navr\'atil}}, \ and\ \bibinfo {author}
  {\bibfnamefont {W.}~\bibnamefont {Nazarewicz}},\ }\href {\doibase
  10.1103/PhysRevC.91.051301} {\bibfield  {journal} {\bibinfo  {journal} {Phys.
  Rev. C}\ }\textbf {\bibinfo {volume} {91}},\ \bibinfo {pages} {051301}
  (\bibinfo {year} {2015})}\BibitemShut {NoStop}%
\bibitem [{\citenamefont {Porro}(2020)}]{PorroMaster}%
  \BibitemOpen
  \bibfield  {author} {\bibinfo {author} {\bibfnamefont {A.}~\bibnamefont
  {Porro}},\ }\href@noop {} {\enquote {\bibinfo {title} {{Importance-Truncation
  techniques in Gorkov-Green’s functions calculations for atomic nuclei}},}\
  } (\bibinfo {year} {2020}),\ \bibinfo {note} {{Master's thesis, Universit\`a
  degli Studi di Milano}}\BibitemShut {NoStop}%
\bibitem [{\citenamefont {Som\`a}\ \emph
  {et~al.}(2020{\natexlab{b}})\citenamefont {Som\`a}, \citenamefont {Barbieri},
  \citenamefont {Duguet},\ and\ \citenamefont {Navr\'atil}}]{Soma21}%
  \BibitemOpen
  \bibfield  {author} {\bibinfo {author} {\bibfnamefont {V.}~\bibnamefont
  {Som\`a}}, \bibinfo {author} {\bibfnamefont {C.}~\bibnamefont {Barbieri}},
  \bibinfo {author} {\bibfnamefont {T.}~\bibnamefont {Duguet}}, \ and\ \bibinfo
  {author} {\bibfnamefont {P.}~\bibnamefont {Navr\'atil}},\ }\href@noop {} {\
  (\bibinfo {year} {2020}{\natexlab{b}})},\ \Eprint
  {http://arxiv.org/abs/2009.01829} {arXiv:2009.01829 [nucl-th]} \BibitemShut
  {NoStop}%
\bibitem [{\citenamefont {Buenker}\ and\ \citenamefont
  {Peyerimhoff}(1975)}]{Buenker75}%
  \BibitemOpen
  \bibfield  {author} {\bibinfo {author} {\bibfnamefont {R.}~\bibnamefont
  {Buenker}}\ and\ \bibinfo {author} {\bibfnamefont {S.}~\bibnamefont
  {Peyerimhoff}},\ }\href {\doibase 10.1007/BF00555301} {\bibfield  {journal}
  {\bibinfo  {journal} {Theoret. Chim. Acta}\ }\textbf {\bibinfo {volume}
  {39}},\ \bibinfo {pages} {217} (\bibinfo {year} {1975})}\BibitemShut
  {NoStop}%
\bibitem [{\citenamefont {Hammer}\ \emph {et~al.}(2020)\citenamefont {Hammer},
  \citenamefont {K\"onig},\ and\ \citenamefont {van Kolck}}]{Hammer20}%
  \BibitemOpen
  \bibfield  {author} {\bibinfo {author} {\bibfnamefont {H.-W.}\ \bibnamefont
  {Hammer}}, \bibinfo {author} {\bibfnamefont {S.}~\bibnamefont {K\"onig}}, \
  and\ \bibinfo {author} {\bibfnamefont {U.}~\bibnamefont {van Kolck}},\ }\href
  {\doibase 10.1103/RevModPhys.92.025004} {\bibfield  {journal} {\bibinfo
  {journal} {Rev. Mod. Phys.}\ }\textbf {\bibinfo {volume} {92}},\ \bibinfo
  {pages} {025004} (\bibinfo {year} {2020})}\BibitemShut {NoStop}%
\bibitem [{\citenamefont {Hjorth-Jensen}\ \emph {et~al.}(2017)\citenamefont
  {Hjorth-Jensen}, \citenamefont {Lombardo},\ and\ \citenamefont {van
  Kolck}}]{CompNucl2017}%
  \BibitemOpen
  \bibinfo {editor} {\bibfnamefont {M.}~\bibnamefont {Hjorth-Jensen}}, \bibinfo
  {editor} {\bibfnamefont {M.~P.}\ \bibnamefont {Lombardo}}, \ and\ \bibinfo
  {editor} {\bibfnamefont {U.}~\bibnamefont {van Kolck}},\ eds.,\ \href
  {\doibase 10.1007/978-3-319-53336-0} {\emph {\bibinfo {title} {An Advanced
  Course in Computational Nuclear Physics}}}\ (\bibinfo  {publisher} {Springer
  International Publishing},\ \bibinfo {year} {2017})\BibitemShut {NoStop}%
\bibitem [{\citenamefont {Brueckner}(1955)}]{Brueckner1955}%
  \BibitemOpen
  \bibfield  {author} {\bibinfo {author} {\bibfnamefont {K.~A.}\ \bibnamefont
  {Brueckner}},\ }\href {\doibase 10.1103/PhysRev.100.36} {\bibfield  {journal}
  {\bibinfo  {journal} {Phys. Rev.}\ }\textbf {\bibinfo {volume} {100}},\
  \bibinfo {pages} {36} (\bibinfo {year} {1955})}\BibitemShut {NoStop}%
\bibitem [{\citenamefont {Goldstone}(1957)}]{Goldstone1957}%
  \BibitemOpen
  \bibfield  {author} {\bibinfo {author} {\bibfnamefont {J.}~\bibnamefont
  {Goldstone}},\ }\href {\doibase 10.1098/rspa.1957.0037} {\bibfield  {journal}
  {\bibinfo  {journal} {Proceedings of the Royal Society of London. Series A.
  Mathematical and Physical Sciences}\ }\textbf {\bibinfo {volume} {239}},\
  \bibinfo {pages} {267} (\bibinfo {year} {1957})}\BibitemShut {NoStop}%
\bibitem [{\citenamefont {Feynman}(1939)}]{Feynman39}%
  \BibitemOpen
  \bibfield  {author} {\bibinfo {author} {\bibfnamefont {R.~P.}\ \bibnamefont
  {Feynman}},\ }\href {\doibase 10.1103/PhysRev.56.340} {\bibfield  {journal}
  {\bibinfo  {journal} {Phys. Rev.}\ }\textbf {\bibinfo {volume} {56}},\
  \bibinfo {pages} {340} (\bibinfo {year} {1939})}\BibitemShut {NoStop}%
\bibitem [{\citenamefont {Nozi{\`e}res}(1963)}]{nozieres}%
  \BibitemOpen
  \bibfield  {author} {\bibinfo {author} {\bibfnamefont {P.}~\bibnamefont
  {Nozi{\`e}res}},\ }\href {https://books.google.fr/books?id=EsPQAAAAMAAJ}
  {\emph {\bibinfo {title} {Le probl{\`e}me {\`a} N corps}}}\ (\bibinfo
  {publisher} {Dunod},\ \bibinfo {year} {1963})\BibitemShut {NoStop}%
\bibitem [{\citenamefont {Kohn}\ and\ \citenamefont
  {Luttinger}(1960)}]{LuttingerWard1}%
  \BibitemOpen
  \bibfield  {author} {\bibinfo {author} {\bibfnamefont {W.}~\bibnamefont
  {Kohn}}\ and\ \bibinfo {author} {\bibfnamefont {J.~M.}\ \bibnamefont
  {Luttinger}},\ }\href {\doibase 10.1103/PhysRev.118.41} {\bibfield  {journal}
  {\bibinfo  {journal} {Phys. Rev.}\ }\textbf {\bibinfo {volume} {118}},\
  \bibinfo {pages} {41} (\bibinfo {year} {1960})}\BibitemShut {NoStop}%
\bibitem [{\citenamefont {Luttinger}\ and\ \citenamefont
  {Ward}(1960)}]{LuttingerWard2}%
  \BibitemOpen
  \bibfield  {author} {\bibinfo {author} {\bibfnamefont {J.~M.}\ \bibnamefont
  {Luttinger}}\ and\ \bibinfo {author} {\bibfnamefont {J.~C.}\ \bibnamefont
  {Ward}},\ }\href {\doibase 10.1103/PhysRev.118.1417} {\bibfield  {journal}
  {\bibinfo  {journal} {Phys. Rev.}\ }\textbf {\bibinfo {volume} {118}},\
  \bibinfo {pages} {1417} (\bibinfo {year} {1960})}\BibitemShut {NoStop}%
\bibitem [{\citenamefont {Dahlen}\ and\ \citenamefont {von
  Barth}(2004)}]{dahlen2004}%
  \BibitemOpen
  \bibfield  {author} {\bibinfo {author} {\bibfnamefont {N.~E.}\ \bibnamefont
  {Dahlen}}\ and\ \bibinfo {author} {\bibfnamefont {U.}~\bibnamefont {von
  Barth}},\ }\href {\doibase 10.1063/1.1650307} {\bibfield  {journal} {\bibinfo
   {journal} {J. Chem. Phys.}\ }\textbf {\bibinfo {volume} {120}},\ \bibinfo
  {pages} {6826} (\bibinfo {year} {2004})}\BibitemShut {NoStop}%
\end{thebibliography}%

\end{document}